\documentclass[11pt,a4paper]{article}
\usepackage{amssymb}
\usepackage{amsmath}
\usepackage[english]{babel}
\usepackage{amsfonts}
\usepackage{rotating}
\usepackage{geometry}
\usepackage{graphicx}
\usepackage{graphicx,color,graphics,lscape,srcltx}
\usepackage{colordvi}
\usepackage{bm,bbm}
\usepackage{amsbsy}
\usepackage{scalefnt, comment}
\usepackage{lscape}
\usepackage{natbib}
\usepackage{subfig}
\usepackage{float}

\providecommand{\U}[1]{\protect \rule{.1in}{.1in}}

\begin{document}

\title{\textbf{Modeling temporal treatment effects with zero inflated
semi-parametric regression models: the case of local development policies in
France}}
\author{Herv\'{e} CARDOT \\
Institut de Math\'{e}matiques de Bourgogne, UMR CNRS 5584, \\
Universit\'{e} de Bourgogne Franche-Comt\'{e} \\
\smallskip \\
Antonio MUSOLESI \\
Department of Economics and Management (DEM), \\
University of Ferrara and SEEDS \\
}
\maketitle

\begin{abstract}
A semi-parametric approach is proposed to estimate the variation along time
of the effects of two distinct public policies that were devoted to boost rural
development in France over the same period of time. At a micro data level, it is often observed that the dependent variable, such as local employment, does not vary along time, so  that we face a kind of zero inflated phenomenon that cannot be dealt with a continuous response model. We introduce a mixture model which
combines a mass at zero and a continuous response. The suggested zero inflated semi-parametric statistical approach   relies on the flexibility
and modularity of additive models with the ability of panel data to deal
with selection bias and to allow for the estimation of dynamic treatment
effects.  In this multiple treatment analysis, we find evidence of
interesting patterns of temporal treatment effects with relevant nonlinear
policy effects. The adopted semi-parametric modeling also offers the
possibility of making a counterfactual analysis at an individual level. The
methodology is illustrated and compared with parametric linear approaches on
a few municipalities for which the mean evolution of the potential outcomes
is estimated under the different possible treatments.
\end{abstract}

\noindent \textit{JEL classification: C14; C23; C54; O18.}

\noindent \textit{Keywords}: Additive Models; Semi-parametric Regression; Mixture of Distributions; Panel Data; Policy Evaluation; Temporal Effects; Multiple Treatments; Local
Development.

\newpage

\section{Introduction}

In response to the deteriorating conditions of distressed areas, many
countries, such as USA, UK and France, have established enterprise zone
programs (EZ) aimed to increase socio-economic development by means of
boosting local employment. At a supranational level, territorial cohesion,
convergence and a harmonious development across regions are among the
objectives of the European Union which tries to pursue through the
structural funds (SF).

Despite their appeal and the high amount of financial resources used, such
geographically targeted policies have been criticized with respect to
different aspects and doubts have been cast with respect to their
effectiveness. As far as EZ are concerned, there exists a number of
micro-econometrics works aiming at assessing their economic effects, which
provide mixed results (for surveys, see \textit{e.g.} %
\citealp{GobillonMagnac2012,PetersFisher2004}). Looking at the analyses of
the effects of regional policies implemented through the European SF, it can
be noted that some earlier studies have been carried out by analyzing the
convergence process and interpreted the descriptive fact of an increasing
divergence across the European regions as an indication that the SF have
been ineffective. More recently, some works adopting a causal framework
appeared \citep{becker2010going,mohl2010eu}, but also for these policies
they provided mixed evidence. In summary, the effectiveness of both EZ and
SF is a relevant and contentious issue in the debate regarding local
development.

We focus on assessing the effect of the EZ and the SF that were devoted
to boost rural development in France. The municipalities, which correspond
to the finest available spatial level, are the statistical units of the
analysis and the dependent variable is the number of employees, as both
programs aim to stimulate employment. The data cover a ten years period,
1993-2002 and such a longitudinal structure constitutes an important source
of identification. Indeed, panel data models have been shown to be very
useful for policy evaluation, allowing to account both for selection on
observables and selection on unobservables, 
and permitting to specify the models in terms of potential outcome at
different points in time %
\citep{HeckmanHotz,HLS99,Wooldridge2005,HsiaoChing2011,Lechner2015}, 
time being an essential element in the notion of causality (\textit{e.g.} %
\citealp{Lechner2011}). Moreover, despite the fact that there is an
increasing availability of relatively long panel data, most of the existing
micro-level studies on regional policies focus on static effects. There are
some exceptions, suggesting that taking account of dynamic effects is
important (see \textit{e.g.} \citealp{Keefe2004,becker2010going}). 

This work provides a new contribution to the literature on regional policy
evaluation revealing for the first time some non-linearities as well as
heterogeneous policy effects that have relevant implications for public
policy design. The paper also introduces methodological advances, allowing
the estimation in a flexible manner of causal effects that can vary over
time and across units. Such an approach could be useful for future research
and outside this specific field of application.

First, it is often observed at a micro data level that the dependent
variable, local employment in the present study, does not vary over time.
This means that when modeling its variations along time we face a kind of
zero inflated phenomenon that cannot be dealt with a continuous response
model. We thus allow the dependent variable to remain constant in time with
a probability that can be strictly larger than zero. To deal with that
phenomenon, a mixture model (see \cite{Mclachlan2000} for a seminal
reference on mixture models) that combines a Dirac mass at zero and a
continuous density is considered.


Second, while a common practice in this literature consists at adopting
parametric models and focusing attention on the mean effect or imposing a
homogeneous effect across units, we relax the parametric specification to model the regression function. 
Specifically, the consideration of a model in which the effect of the policy
is expanded as a nonparametric function of some variables provides a richer
framework that allows for a refined analysis at an individual level and
permits to highlight heterogeneous policy effects, which are missed when
focusing on mean effects. We rely on the rather general framework of
additive models and generalized additive models \citep{MR2206355}, giving
much more flexibility and robustness than usual linear models, but also
addressing the curse of dimensionality problem arising in fully
nonparametric models, which could be an extremely serious problem because of
the large number of potential regressors. Penalized splines are used to
represent the non parametric parts of the additive model %
\citep{MR2090902,wood2008fast} as they have been proven to be useful
empirically in many aspects (see, e.g. \citealp{ruppert03}) and, in recent
years, their asymptotic properties have been studied and then connected to
those of regression splines, to those of smoothing splines and to the
Nadaraya - Watson kernel estimators (see, \textit{e.g.} %
\citealp{li2008asymptotics}, \citealp{Wood2016}). The estimation is finally
carried out by maximizing the corresponding likelihood function, which is a
mixture of a mass at zero and a continuous density.



Finally, the proposed semi-parametric modeling also permits to estimate what
would have been the expected effects of such policies on particular
municipalities by performing a counterfactual estimation at an individual
level. The evolutions of the potential outcomes are thus estimated and
compared under the different possible treatments for a few municipalities.
These municipalities, selected with a clustering $k$-medoids algorithm (see %
\citealp{KaufmanRousseeuw1990}), represent communes with different but
typical characteristics within their cluster. A comparison of the results with those obtained from  some standard parametric continuous response models finally provides interesting insights into the size of the bias that may arise when a parametric specification is imposed  or the mass of observations at zero is not accounted for.

It is also worth noting that while most of the previous studies focus on one
particular policy, either EZ or SF, we will assess the effect of both
policies as well as their interaction by adopting a multiple treatments
framework (see \citealp{Frolich2004}, for a survey).

The remainder of the paper is structured as follows. Section two describes
the rural policies adopted in France, presents the data and provides some
descriptive statistics. Section three is devoted to the presentation of the
econometric framework and of the estimation methodology. Section four
provides the presentation and discussion of our main results while section
five summarizes and concludes. Additional results are given in a
supplementary file. 

\section{Description of the policies and data}

In France, EZ have been implemented to boost job creation. Such policies are
based on fiscal incentives to firms located in deprived areas. Specifically
designed to boost employment of rural areas, the ZRR (\textit{Zones de
Revitalisation Rurale) }program started the 1st September 1996. A noticeable
feature of the program is that the selection of ZRR was clearly not random.
A rather complex algorithm was used to determine the eligibility, according
to some observable -- demographic, economic and institutional -- criteria.
To be eligible to ZRR, a municipality should be a part of a canton with
population density lower than 31 inhabitants per square km (1990 Population
Census)\footnote{%
A canton with a population density less than 5 inhabitants per square km is
automatically labelled as ZRR without any other requirement.}. The
population or the labor force must also have diminished or the share of the
agricultural labor employment must be at least twice the French average.
Finally, to be included into the program, the municipality should belong to
a pre-existing zoning scheme set up by the European Union, which is called
TRDP (\textit{Territoire Rural de D\'{e}veloppement Prioritaire}). However,
due to political tempering, it is also likely that, beyond such observed
criteria, other sources of selection on unobservables could affect the
process (\citealp{GobillonMagnac2012}). A more detailed description of the
ZRR program can be found in \cite{BehaghelLorenceau2015}. 

Beyond the French experience, EZ have been largely criticized with respect
to several aspects, such as the possibility of i) windfall effects to firms
who would have hired workers even in absence of the policy; ii) negative
spatial spillovers because EZ does not necessarily result in job creation
but could cause geographical shifts in jobs from non-EZ to EZ areas; iii)
stigmatization of the targeted neighborhood; iv) in absence of tax revenue
compensation, EZ could lead to a decrease in the local provision of public
services and v) obtaining only a transitory effect on employment and the
need for integrated policies against structural unemployment.

At a supranational level, the SF are addressed to help lagging or
re-structuring regions, so they are given to regions upon their economic
characteristics (such as the per capita GDP or the unemployment level) and
then are assigned from the regions to firms or to public actors (top-down
process) without a clearly expressed assignment mechanism. Then, \ also for
these policies, sources of selection on both observables and unobservables
are expected to be relevant. Specifically devoted to boost rural
development, the objective 5B programs (1991-93 and 1994-99) allocated
financial subsides to firms and public actors located in eligible \textit{%
\textquotedblleft rural areas in decline\textquotedblright }. The
eligibility criteria for belonging to an objective 5B area (canton) required
that the area has a high share of agricultural employment, a low farming
income and a low level of per capita GDP (Gross Domestic Product). The main
goal of 5B programs was to improve economic development and local
infrastructures, and to support the activities of farms, small and medium
sized firms, rural tourism. 

Our sample is obtained by merging different data sets. The municipalities,
which correspond to the finest available spatial level, are the statistical
units of the analysis and the dependent variable is the number of employees.
The data were obtained over a period of ten years, 1993-2002 (for each year
data refer to the 1st January), from the INSEE (\textit{Institut National de
la Statistique et des Etudes Economiques}) and SIRENE (\textit{Syst\`{e}me
Informatique pour le R\'{e}pertoire des Entreprises et de leurs \'{E}%
tablissements}) sheet. As explanatory variables, we dispose of ZRR zoning
during the period and of the 5B zoning over the period 1994-99. Some other
explanatory variables come from the CENSUS. Since the CENSUS data are
collected every ten years, and in order to control for the initial
conditions, we use data from 1990 CENSUS. Such CENSUS data have been
provided by the INSEE in separate sheets, gathering demographic, education
and work's qualification information. Finally, we also have at hand
information on land use in 1990, obtained thanks to satellite images. After
the merging process and some cleanings that are detailed in Appendix A of
the supplementary file we obtain a sample of 25593 municipalities.

It can be seen in Table~\ref{table:statdesc} that about 30\% of the 25593
municipalities in our sample were under the ZRR scheme. Over the period
1994-99, about 47\% of the municipalities were under objective 5B. Examining
ZRR and 5B jointly, it appears that 50.9\% of the municipalities were under
at least one of the two policies. Only 27.4\% of the municipalities were, in
our sample, under both policies, whereas 20.6\% received a support only from
5B program and 2.8\% of the municipalities received the incentives only from
ZRR. As expected, the treated municipalities present lower socio-economic
performances compared to the non-treated ones, with the municipalities under
objective 5B alone performing generally better than the other treated
municipalities. Also note that for the estimation of treatment effects, the
only partial overlap between ZRR and 5B programs is a useful source of
identification, which is exploited in this paper to estimate the specific
effect of each policy as well as their interaction effect.

\section{Model specification and estimation}


We borrow notations from \cite{HeckmanHotz} and \cite{Frolich2004}. Let $i$
denote a statistical unit (a municipality in our framework) which is
assigned to one of $R$ mutually exclusive development incentives. We denote
by $Y_{it}^{r}$ the potential employment level for municipality $i$ at time $%
t$ under treatment (incentive) $r$, for $r\in \{0,1,\ldots ,R-1\}$, with the
convention that $r=0$ corresponds to no treatment. Time $t$ is discrete,
taking values in $t_{0}<t_{1}<\ldots <t_{m}$. We assume that the incentives
are allocated after $t_{0}$ and that they may produce an effect from period $%
k$, with $t_{k}>t_{0}$. All the counterfactuals are assumed to be equal
before the treatment begins, that is to say $Y_{it}^{r}=Y_{it}^{0}$ for $%
t_{0}\leq t<t_{k}$ and $r=1,2,\ldots ,R-1$. As a starting point, we consider
the following general model, 
\begin{align}
Y_{it}^{r}& =Y_{it}^{0}\qquad \qquad t_{0}\leq t<t_{k},  \notag \\
& =Y_{it}^{0}+\Delta_{it}^{r},\qquad t_{k}\leq t\leq t_{m},
\label{def:modelgene}
\end{align}%
where $Y_{it}^{0}$ is the employment level for municipality $i$ at time $t$
in the absence of development funds ($r=0$). For time $t\geq t_{k}$, $%
\Delta_{it}^{r}$ is simply the difference between $Y_{it}^{r}$ and $%
Y_{it}^{0}$, that is to say the differential effect on the potential
outcome, compared to no treatment at all, of treatment $r$ on unit $i$. With
this general model, $\Delta_{it}^{r}$ is allowed to vary from one
statistical unit to another and also to depend on time $t$.

\label{sec:mixture} Let us denote by $D_{i}$, with $D_{i}\in \{0,1,\ldots
,R-1\}$, the treatment status of municipality $i$, that is supposed to be a
random variable. Consider now a set of characteristics $X_{i}=(X_{i1},\ldots
,X_{ip})$ observed during the first period of time $t_{0}$, which are the 
\textit{initial conditions}.

\subsection{Identification issues}

A classical condition in policy evaluation (see %
\citealp{ImbensWooldridge2009}), generally referred to as conditional
independence assumption, unconfoundedness or selection on observables, is
that 
\begin{equation}
Y_{it}^{r}\perp \!\!\!\perp D_{i}\ |\ X_{i},  \label{CIA-s}
\end{equation}%
so that the information contained in the observed variables $X_{i}$ makes
the potential outcomes $Y_{it}^{r}$ unconfounded, that is, conditionally
independent of the treatment status $D_{i}$ given $X_{i}$.

Since selection bias may not be completely eliminated even after controlling
for the observables $X_{i}$, it is also important to note that a \textit{%
before-after} approach may help to address the issue of selection on
unobservables. We thus consider that the conditional independence assumption
(\ref{CIA-s}) holds for the difference of the outcome after and before the
beginning of the policy, 
\begin{equation}
Y_{it}^{r}-Y_{it_{0}}^{0}\perp \!\!\!\perp D_{i}\ |\ X_{i},\qquad \forall
r\in \{0,1,\ldots ,R-1\}.  \label{CIA_B}
\end{equation}%
The new conditional independence assumption (\ref{CIA_B}) is a less
restrictive condition than (\ref{CIA-s}).\footnote{%
See Appendix B for a more detailed discussion.}

It is worth mentioning that we could consider propensity scores (\citealp
{RosenbaumRubin1983,angrist2004control,imai2004causal}) in place of $X$, in
the conditioning variables appearing in (\ref{CIA_B}). This would ensure
that $D$ is conditionally independent of the potential outcomes while
achieving dimensional reduction. One drawback of this approach, which can be
effective for estimating mean effects on the treated or on the whole
population, is interpretation (see e.g. \citealp {ImbensWooldridge2009}) as
well as the fact that the propensity scores may not be highly relevant
variables to estimate accurately the variations of the conditional potential
outcomes, given the vector of covariates $X$. Indeed, we can split the
vector of all the available covariates $X$ into four parts, 
\begin{equation*}
X=\left( X_{Y\cap D},X_{\bar{Y}\cap D},X_{Y\cap \bar{D}},X_{\bar{Y}\cap \bar{%
D}}\right) ,
\end{equation*}%
where $X_{Y\cap D}$ is the set of covariates that are related both to $%
Y_{t}^{r}-Y_{t_{0}}^{0}$ and $D$ and $X_{\bar{Y}\cap D}$ is the set of
covariates that are independent of $Y_{t}^{r}-Y_{t_{0}}^{0}$ but are related
to $D.$ Note that these two sets, $X_{Y\cap D}$ and $X_{\bar{Y}\cap D}$,
represent the variables entering the propensity score function. The set $%
X_{Y\cap \bar{D}}$ is the set of covariates that are related to $%
Y_{t}^{r}-Y_{t_{0}}^{0}$ but are independent of $D$ and $X_{\bar{Y}\cap \bar{%
D}}$ is the set of covariates that are independent of $%
Y_{t}^{r}-Y_{t_{0}}^{0}$ and $D$ (see Figure \ref{fig:causality}).

\begin{center}
\textbf{Figure \ref{fig:causality} about here}
\end{center}

The smallest set of conditioning variables required to satisfy condition (%
\ref{CIA_B}) is $X_{Y\cap D}$. However, since one of the aims in this work is to
estimate, at an individual level, the variation over time of the expected
potential effects of the different policies, we also take
account of the set of variables $X_{Y\cap \bar{D}}$ in a way that is as
flexible as possible to have a better prediction of the potential outcomes.
As a result, our statistical approach is built by modeling in a non
parametric way the relation between $Y_{t}^{r}-Y_{t_{0}}^{0}$ and $X$ and by
selecting, among all the available variables, the variables that belong to
one of the two sets $X_{Y\cap D}$ and $X_{Y\cap \bar{D}}$. Note that if we
were interested in the best possible estimation of the propensity scores, i{%
.e.} the scores giving the probability of receiving policy $r$, for $%
r=0,\ldots ,R-1$, our statistical models would have focused on the sets of
variables $X_{\bar{Y}\cap D}$ and $X_{Y\cap D}$.

In the following Sections it is assumed that the set of covariates $X$ is
restricted to $X_{Y\cap D}$ and $X_{Y\cap \bar{D}}$. Other observed
variables that could be considered are those that influence selection into
the program even if they do not affect directly the outcome, \textit{i.e.} $%
X_{\bar{Y}\cap D}$. Introducing these variables in the regression function
may help to solve the problem of selection on observables, provided there is
no misspecification error, using the terminology by \cite{HeckmanHotz}.
Appendix A provides further comments on this issue while the variable selection
procedure is described in Section~\ref{sec:Results}.

\subsection{Zero inflation and econometric modeling}

\begin{figure}[H]
\includegraphics[width=150mm]{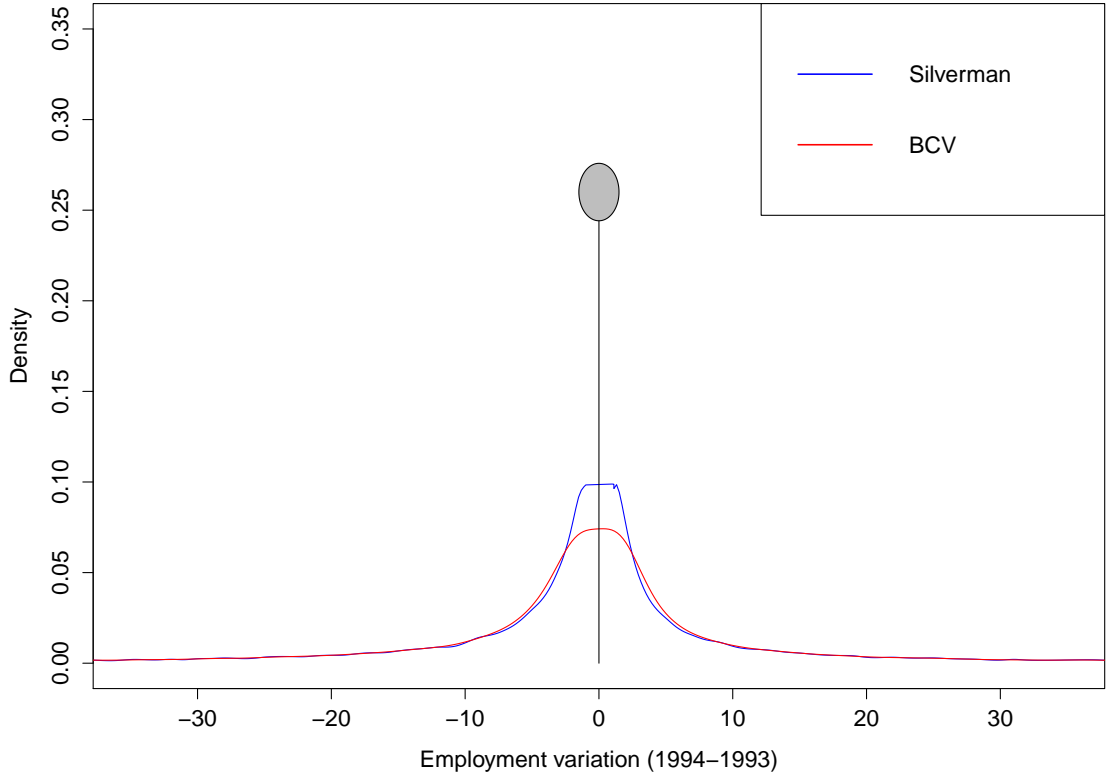}
\caption{The estimated distribution of $Y_{it}^{D_{i}}-Y_{it_{0}}$ for $%
t=1994$ and $t_{0}=1993$. The probability of observing no variation is
estimated by the proportion of observations such that $%
Y_{it}^{D_{i}}-Y_{it_{0}}=0$ whereas the continuous density of $%
Y_{it}^{D_{i}}-Y_{it_{0}}\neq 0$ is estimated thanks to kernel density
estimators, with two different standard ways of selecting the bandwidth
value. Silverman: Silverman's rule of thumb; BCV: Biased Cross Validation
(see \citealp{Silverman86,sheather2004density}). }
\label{Density}
\end{figure}

A relevant feature of this study is that the statistical units are generally
demographically small and we observe no variation at all of the dependent
variable along time for a non negligible fraction of the municipalities, 
\textit{i.e.} $Y_{it}^{D_{i}}=Y_{it_{0}}$. Table~\ref{table_desc2} in the
Supplementary file shows that the modal value of $Y_{it}^{D_{i}}-Y_{it_{0}}$
is indeed $0$ for all the values of $t$, with $t$ varying between 1994 to
2002 and $t_{0}$ corresponding to the year 1993. We can also remark that the
fraction of zeros decreases with $t$ and varies with the treatment status.
The estimated distribution of the dependent variable, $%
Y_{it}^{D_{i}}-Y_{it_{0}}$, for $t=1994$, which is a mixture of a mass at 0
and a continuous density function, is depicted in Figure~\ref{Density}.

\begin{center}
\textbf{Figure \ref{Density} about here}
\end{center}

This empirical fact leads us to introduce a new econometric model that is
able to take account of this important feature of the data. There is a kind
of \textit{zero inflated effect} that can not be dealt with a classical
continuous response model. We thus allow $Y_{it}^{D_{i}}-Y_{it_{0}}^{0}$ to
be equal to zero with a probability that may be strictly larger than $0$.
Let us denote by $\Delta _{it}^{D_{i}}=\sum_{r=0}^{R-1}{\mathbbm1}%
_{\{D_{i}=r\}}\Delta _{it}^{r}$, with $\Delta
_{it}^{r}=Y_{it}^{r}-Y_{it_{0}}^{0}$. We propose to describe the
distribution of the counterfactual variation of the level of employment $%
\Delta _{it}^{r}$ as a mixture of a mass at $0$ and a continuous
distribution. Using the decomposition $Y_{it}^{r}-Y_{it}^0 =
Y_{it}^{r}-Y_{it_{0}}^0 -\left(Y_{it}^{0}-Y_{it_{0}}^0\right)$, we obtain
that the expected conditional effect at time $t$ of policy $r$ compared to
no policy is expressed as follows, 
\begin{align}
\mathbb{E}\left[ Y_{it}^{r}-Y_{it}^0 \mid X_{i}\right] &= \mathbb{E}\left[
Y_{it}^{r}-Y_{it_0}^0 \mid X_{i}\right] - \mathbb{E}\left[
Y_{it}^{0}-Y_{it_0}^0 \mid X_{i}\right]  \notag \\
& = \mathbb{E}\left[ Y_{it}^{r}-Y_{it_0}^0 \mid X_{i}, \Delta _{it}^{r} \neq
0 \right] \times \left( 1-\mathbb{P}\left[ \Delta_{it}^{r}=0\mid X_{i}\right]
\right)  \label{eq:deltar} \\
& \ \ \ - \mathbb{E}\left[ Y_{it}^{0}-Y_{it_0}^0 \mid X_{i}, \Delta
_{it}^{0} \neq 0 \right] \times \left( 1-\mathbb{P}\left[ \Delta_{it}^{0}=0%
\mid X_{i}\right] \right)  \label{eq:delta0}
\end{align}

Expressions (\ref{eq:deltar}) and (\ref{eq:delta0}), which
explicitly take account of the zero inflation feature of the counterfactual
outcome variations, are the main object of interest in this paper.

\subsection{A flexible semi-parametric modeling approach}

\label{sec:gam}

Suppose now we have a sample $(Y_{it}^{D_{i}},X_{i},D_{i})_{i=1,\ldots ,n}$,
for $t\in \{t_{0},\ldots ,t_{m}\}$. We can write 
\begin{equation}
Y_{it}^{D_{i}}=\sum_{r=0}^{R-1}Y_{it}^{r}{\mathbbm1}_{\{D_{i}=r\}}
\label{decomp:y}
\end{equation}%
where the indicator function satisfies ${\mathbbm1}_{\{D_{i}=r\}}=1$ if $%
D_{i}=r$ and zero else. Consequently, we can express the expected variation
along time $Y_{it}^{D_{i}}-Y_{it_{0}}^{0}$, of the employment level of
municipality $i$ given that $\Delta_{it}^{D_i} \neq 0$, as follows, 
\begin{align}
\mathbb{E} \left[ Y_{it}^{D_{i}}-Y_{it_{0}}^{0} | X_i, \Delta_{it}^{D_i}
\neq 0 \right] &= \mu_t^0(X_i) + \sum_{r=1}^{R-1}{\mathbbm1}_{\{D_{i}=r\}}
\alpha_{t}^{r}(X_i),  \label{model:diffsample}
\end{align}
The term $\alpha _{t}^{r}(X_i)$ which reflects in (\ref{model:diffsample})
the impact of treatment $r$ should be equal to zero when $t_{0}\leq t<t_{k}$
whereas $\mu_t^0(X_i)$ corresponds to the expected variation under no policy.

Introducing  (\ref{model:diffsample}) in  (\ref{eq:deltar}) and (\ref{eq:delta0}), we can also express, given $X_i$, the expected effect of policy $r$ at time $t$ as follows
\begin{align}
\mathbb{E}\left[ Y_{it}^{r}-Y_{it}^0 \mid X_{i}\right] &= \left( 1-\mathbb{P}\left[ \Delta _{it}^{r}=0\mid X_{i}\right]
\right) \times \alpha_{t}^{r}(X_i)   \nonumber \\
 & -\left( \mathbb{P}\left[ \Delta _{it}^{r}=0\mid X_{i}\right]-\mathbb{P} \left[ \Delta _{it}^{0}=0\mid X_{i}\right] \right) \times \mu_{t}^{0}(X_{i}).
\label{def:effectzero} 
\end{align}
The conditional expected counterfactual in (\ref{def:effectzero}) is composed of two main terms that may
act in opposite directions, so that interpretation is more difficult
compared to usual policy evaluation models based on continuous response regression models  that
do not take account of the zero inflation effect.

In the econometric literature, a common practice consists in modeling $\mu
_{t}^{0}(X_{i})$ and $\alpha _{t}^{r}(X_{i})$ using parametric
specifications, where the $\mu _{t}^{0}(X_{i})$ is usually a linear
function, $\mu _{t}^{0}(X_{i})=\mu _{t}^{0}+\sum_{j=1}^{p}\beta
_{jt}^{0}X_{ij}$, and the term $\alpha _{t}^{r}(X_{i})$  does vary
with the covariates or  is expanded as a linear function of them
(see \textit{e.g.} \citealp{HeckmanHotz}, eq. 3.9). The linearity assumption
is strong and a miss-specification of the relation between $%
Y_{it}^{r}-Y_{it_{0}}$ and the regressors may lead to wrong results and
interpretation of the policy effect. We thus prefer to consider a more
general model that can take account of non linear effects nonparametrically
via an additive form \citep{MR1082147,MR2206355}. This also makes
the underlying identification conditions less restrictive \citep{lechner2011estimation}.

The expected value that would be obtained at time $t$ for a municipality
with characteristics $X_{i}$ under no treatment, is supposed to be
additively modeled as follows, 
\begin{equation}
\mu_t^0(X_{i})=\mu _{t}^{0}+\sum_{j=1}^{p}g_{jt}^{0}(X_{ij}),
\label{def:amy0}
\end{equation}%
where $g_{jt}^{0}(.),\ j=1,\ldots ,p$, are unknown smooth univariate
functions. The identifiability constraints 
\begin{equation*}
\mathbb{E} \left[  g_{jt}^{0}(X_{j}) \ | \ \Delta_{t}^0 \neq 0 \right] = 0, \quad j=1,\ldots ,p,   
\end{equation*}%
ensure that $\mu _{t}^{0}$ represents the mean value of the variation of the
potential outcome between $t$ and $t_{0}$ if all the units in the population would have
received no incentives at all.

A key assumption of this paper is that the conditional differential policy
effect $\alpha _{it}^{r}=\alpha _{t}^{r}(X_{i})$ can be expressed, given the
vector of covariates $X_{i}$, with the following additive model, 
\begin{align}
\alpha _{it}^{r} &=\alpha _{t}^{r}+\sum_{j=1}^{p}g_{jt}^{r}(X_{ij}),
\label{def:alpharAM}
\end{align}%
where $g_{jt}^{r}(.),\ j=1,\ldots ,p$ are unknown smooth functions
satisfying the identifiability constraints 
\begin{equation*}
\mathbb{E} \left[ g_{jt}^{r}(X_{j}) | \ \Delta_{t}^r \neq 0 \right]   =0, \quad j=1,\ldots ,p.
\end{equation*}%
Consequently, $\alpha _{t}^{r}$ represents the mean effect, over the whole population, at period $t$ of treatment $r$ and the function $g_{jt}^{r}(.)$ reveals
how the mean impact of the policy $r$ is modulated by the individual characteristics of each considered statistical unit.

Note that a simple extension of (\ref{def:alpharAM}) consists in considering
interactions between covariates instead of additive effects. For $2\leq
d\leq p$, the additive effects of $d$ covariates, $%
g_{1t}^{r}(X_{i1})+g_{2t}^{r}(X_{i2})+\cdots +g_{dt}(X_{id})$ can be
replaced by a more general multivariate function
\begin{align*}
\alpha _{it}^{r} &=\alpha _{t}^{r}+ g_{1,2,\ldots ,d,t}^{r}(X_{i1},X_{i2},\ldots ,X_{id})
\end{align*}%
that could allow a more flexible fit to the data, at the expense of a more
difficult interpretability and, because of the curse of dimensionality, less
precise estimates. 
The behavior of functions $g_{jt}^{r}$ is of central interest and our
general model encompasses the following particular cases, i) no effect of
the policy $r$ compared to no treatment at all, when $\alpha _{t}^{r}=0$ and 
$g_{jt}^{r}=0$ for all $t\geq t_{k}$; 
 ii) linear trends in time when $\alpha _{t}^{r}=\alpha _{0}^{r}+\alpha
_{1}^{r}t$ and linear effects of the covariates when $g_{jt}^{r}(X_{ij})=%
\beta _{jt}^{r}X_{ij}$ and iii) polynomial trends in time and polynomial
effects of the covariates, as well as smooth threshold effects.

\medskip

We suppose that the probability that $Y_{it}^{D_{i}}-Y_{it_{0}}=0$ given the
covariates can be expressed with a generalized additive model and a logit
link function. Using a similar decomposition as in~(\ref{decomp:y}), we
consider the following logistic regression models, for $t=t_{1},\ldots
,t_{m} $, 
\begin{equation}
\mbox{logit}\left( \mathbb{P}\left[ Y_{it}^{D_{i}}-Y_{it_{0}}^{0}=0\mid X_{i}%
\right] \right) =\beta _{0t}^{0}+\sum_{r=1}^{R-1}{\mathbbm1}%
_{\{D_{i}=r\}}\delta \beta _{0t}^{r}+\sum_{j=1}^{p}\beta _{jt}(X_{ij}),
\label{def:gamlogit}
\end{equation}%
where $\beta _{jt}(.)$ are unknown smooth univariate functions. For our
purpose, the most important parameters are the differential effects $\delta
\beta _{0t}^{r}$, $r=1,\ldots ,R-1$. For example, if $\delta \beta
_{0t}^{r}>0$, then the probability no variation is larger under policy $r$
compared to no policy at all ($r=0$) given the covariates $%
X_{i}=(X_{i1},\ldots ,X_{ip})$. Recall that the unknown functions $\beta
_{jt}(X_{ij})$ are not necessarily linear and that it would be possible to
consider a more sophisticated model that could take interaction effects into
account, replacing $\beta _{jt}(X_{ij})$ by $\beta _{jt}^{r}(X_{ij})$, for $%
r=1,\ldots ,R-1$.

\subsection{Estimation procedure}

We observe, for a statistical unit $i$, the realized outcomes $Y_{it}^{{D}%
_{i}}$ at instants $t=t_{0},\ldots ,t_{m}$, whereas the counterfactuals $%
Y_{it}^{r}$, for $r\neq D_{i}$, cannot be observed. The estimation of the
parameters and functions defined in (\ref{def:amy0}), (\ref{def:alpharAM})
and (\ref{def:gamlogit}), relies on the sample $%
(Y_{it}^{D_{i}},X_{i},D_{i})_{i=1,\ldots ,n}$, for $t\in \{t_{0},\ldots
,t_{m}\}$. We assume that there are no spatial interactions between the
statistical units so that $(Y_{it}^{D_{i}},X_{i},D_{i})$ and $(Y_{\ell
t}^{D_{\ell }},X_{\ell },D_{\ell })$ can be supposed to be independent if $%
i\neq \ell $. This hypothesis can be easily relaxed by allowing, for
instance, spatial spillover effects via the definition of additional
covariates that take account of the treatments received by the neighboring
municipalities (see Appendix \ref{sec:spatialspillover}). The $t_{m}-t_{0}$
samples $(Y_{it}^{D_{i}}-Y_{it_{0}}^{0},X_{i},D_{i})_{i=1,\ldots ,n}$, with $%
t=t_{1},\ldots ,t_{m}$ are used separately to estimate the parameters of
interest and the regression functions.

The fact that the considered mixture is a mixture of a continuous variable
and a discrete variable makes the computation of the likelihood rather
simple compared to mixtures of continuous variables or mixtures of discrete
variables (see \cite{Mclachlan2000}). Indeed, as far as the continuous part
is concerned, the probability of no variation is equal to zero and we can
proceed as if the two underlying distributions were adjusted separately.
Assuming $(Y_{it}^{D_{i}}-Y_{it_{0}}^{0},\ i=1\ldots ,n)$ are conditionally
Gaussian and independent random variables, the likelihood at each instant $t$%
, is given by 
\begin{equation*}
\mathcal{L}_{t}=%
\prod_{i=1}^{n}p_{it}^{T_{it}}(1-p_{it})^{1-T_{it}}f_{t}^{D_{i}}(Y_{it}^{D_{i}}-Y_{it_{0}}^{0};X_{i},D_{i})^{1-T_{it}}
\end{equation*}%
where $T_{it}={\mathbbm1}_{\{\Delta _{it}^{D_{i}}=0\}}$ is the indicator
function of no variation between $t$ and $t_{0}$ and $p_{it}=\mathbb{P}\left[
\Delta _{it}^{D_{i}}=0\mid X_{i},D_{i}\right] $. Taking account now of the
different policies, the log-likelihood can be expressed as follows, 
\begin{align}
\ln \mathcal{L}_{t}& =\sum_{i:T_{it}=1}\ln p_{it}+\sum_{i:T_{it}=0}\ln
(1-p_{it})  \label{def:loglik0} \\
& +\sum_{i:T_{it}=0}\sum_{r=0}^{R-1}{\mathbbm1}_{\{D_{i}=r\}}\ln
f_{t}^{r}(Y_{it}^{D_{i}}-Y_{it_{0}}^{0};X_{i},D_{i}),  \label{def:loglik}
\end{align}%
so that the probability of no variation can be estimated separately by
maximizing the terms at the right-hand side of (\ref{def:loglik0}), whereas
the additive models related to the continuous variation of $%
Y_{t}^{D}-Y_{t_{0}}^{0}$ are estimated by maximizing the function at the
right-hand side of (\ref{def:loglik}). This means that in practice, the
subsample $\{i\ |T_{it}=0\}$ is used for the adjustment of the additive
models related to the continuous part. The estimation of the unknown
functional parameters introduced in (\ref{def:amy0}), (\ref{def:alpharAM})
and (\ref{def:gamlogit}), which are supposed to be smooth functions, is
performed thanks to the \texttt{mgcv} library in the R language (see %
\citealp{MR2206355}, for a general presentation). The regression functions
to be estimated are expanded in spline basis and a penalized likelihood
criterion is maximized. Penalties, tuned by smoothing parameters, are added
to the log-likelihood in order to control the trade off between smoothness
of the estimated functions and fidelity to the data. To select the values of
the smoothing parameters, restricted maximum likelihood (REML) estimation
was preferred over alternative approaches such as Generalized Cross
Validation (GCV) or Akaike's Information Criterion (AIC), since such
approaches may lead to under-smoothing and are more likely to develop
multiple minima than REML. Pointwise confidence intervals that take account
of the smoothing parameter uncertainty can be obtained as in \cite{Wood2016}
and variable selection is performed following \cite{MarraWood2011}.

\section{Results}

\label{sec:Results}

The main goal of this paper is the estimation of the mean differential
effect, $\mathbb{E}\left[ Y_{it}^{r}-Y_{it}^{0}\mid X_{i}\right] $, of
policy $r$ compared to no policy, for a unit with characteristics $X_{i}$.
This conditional expectation, which is expressed in~(\ref{def:effectzero}),
depends on different ingredients. Estimations of $\alpha _{t}^{r}(.)$ and $%
\mu _{t}^{0}(.)$ are related to the continuous part of the model while the
discrete one provides us information about the conditional probabilities $%
\mathbb{P}\left[ \Delta _{it}^{r}=0\mid X_{i}\right] $ and $\mathbb{P}\left[
\Delta _{it}^{0}=0\mid X_{i}\right] $.

After a discussion about the parameters of interest, we briefly present some
preliminary estimation results for both the continuous and the discrete part
of the model, taken separately. We then provide our main results. Since our
model allows the policy effect to vary both in time and across units, we
specifically provide a temporal counterfactual analysis at an individual
level, with an illustration on a few representative municipalities for which
the evolutions of the potential outcomes are estimated and compared under
the different possible treatments. This could provide interesting economic
and policy oriented advices. We finally provide some insights into the size
of bias that may arise when a parametric specification is imposed or the
mass of observations at zero is not not accounted for, by comparing the
proposed approach with some standard methods.

\subsection{Parameters of interest}

\label{sec:param}

We focus on the assessment of ZRR and 5B as well as their joint mean effect.
The partial overlap of these two schemes makes possible the identification
of the interaction effect of ZRR and 5B. We thus adopt a framework with $R=4$
multiple potential outcomes and consider the generalized treatment variable, 
\texttt{D}$_{i}\in \left\{ 0,ZRR,5B,ZRR\&5B\right\} $ indicating the
programme in which municipality $i$ actually participated. The modality $0$
indicates that the municipality \textit{i} did not receive any policy, $ZRR$
(respectively $5B$) indicates that the municipality \textit{i} received
incentives only from ZRR (respectively only from 5B) and $ZRR\&5B$ indicates
that the municipality \textit{i} received incentives both from ZRR and 5B.

As far as the continuous response is concerned, the parameter $\alpha
_{t}^{5B}$ measures the mean differential effect, over the whole sample, of
policy 5B compared to no policy at all ($r=0$) whereas the joint effect of
ZRR and 5B is given by $\alpha _{t}^{ZRR\&5B}$. Finally, concerning the
effect of ZRR, it can be noticed that only a few municipalities (precisely
722) are treated in this case. Consequently, we prefer to focus our
attention on the 7014 municipalities that receive incentives both from 5B
and ZRR and we calculate the following differential effect $\alpha
_{t}^{ZRR}=\alpha _{t}^{ZRR\&5B}-\alpha _{t}^{5B}$. This differential effect
simply represents the mean difference between the outcome when receiving
incentives both from ZRR and 5B and the outcome when only 5B applies. The
same reasoning applies to the interpretation of the\ expected conditional
differential effect $\alpha _{it}^{r}$ in (\ref{def:alpharAM}) as well as
for the parameter $\delta \beta _{0t}^{r}$ when dealing with the estimation
of the conditional probability of a null employment variation in (\ref%
{def:gamlogit}).

\subsection{Preliminary results}

\label{sec:preliminary}

\subsubsection*{The continuous response}

\label{sec:continuous}

As far as the continuous response is considered, additive models are fitted
on the subsamples $\{i\ |Y_{it}^{D_{i}}-Y_{it_{0}}\neq 0\}$. We focus
attention on the\ expected conditional differential effect $\alpha _{it}^{r}$%
, for $t_{k}\leq t\leq t_{m}$ and $r\in \{1,\ldots ,R-1\}$, while the
results about the effect of the initial conditions, which enter
nonparametrically via the additive smooth functions $%
\sum_{j=1}^{p}g_{jt}^{0}(X_{ij})$ in (\ref{def:amy0}), are not discussed
here but are available upon request.

A backward variable selection procedure has been employed to select the
variables to be introduced in the regression functions defined in (\ref%
{def:amy0}) such that the conditional independence assumption (\ref{CIA_B})
holds. This procedure leaded us to retain 11 variables among the 16 initial
variables (the selected variables are those reported in Table~\ref%
{table:statdesc} in the Supplementary file).

We consider pre-treatment covariates, say $X_{pre},$ in the set of
observable variables $X$ to ensure that $D$ causes $X$ and $Y$ causes $X$ do
not occur.\footnote{\cite{lee2005micro} labels \textit{collider} the
situation when both $D$ and $Y$ cause $X$.} This is likely to be relevant in
our economic context where it could be expected that the covariates prior
the introduction of the policy, such as for example the share of qualified
workers or the existing stock of infrastructure, cause both the inclusion in
the program $D$, and the potential local employment $Y$ ($X_{pre}\rightarrow
D$ and $X_{pre}\rightarrow Y)$. After the introduction of the policy, the
level of such covariates, say $X_{post}$, is likely to be affected by its
past values $X_{pre}$, by the treatment $D$ and finally also by the response
variable $Y$. Indeed, in the example mentioned above, the share of qualified
workers and the stock of infrastructure may be directly affected by the
policy ($D\rightarrow X_{post}$) and since the introduction of the policy
could have also increased local employment ($D\rightarrow Y$), this may in
turn stimulate the creation of new infrastructure/qualified hires ($%
Y\rightarrow X_{post}$). In such a causal framework, $X_{pre}$ should be
controlled for whereas $X_{post}$ should not (see \citealp{lee2005micro,
lechner2011estimation}).

Also note that the vector $X_{i}$ may contain the initial level of
employment. Including the initial outcome as a regressor is particularly
relevant if the average outcomes of the treated and the control groups
differ substantially at the first period, as in this case (see \textit{e.g.} %
\citealp{ImbensWooldridge2009,Lechner2015})

As expected, the initial outcome was found highly significant and has been
included. In almost all cases the linearity was clearly rejected in favor of
nonlinear regression functions. We \ also remark that not imposing a linear
relation in (\ref{def:amy0}) leads to retain a larger number of significant
variables compared to simpler linear regression models since there are only
6 significant variables when imposing linear relations.\footnote{%
Also note that almost the same results would have been obtained if we would
have employed the double penalty variable selection approach proposed by 
\cite{MarraWood2011} (detailed results are available upon request).}

Since a major interest lies in assessing possible heterogeneous treatment
effects, we examine how the effect of a policy may vary with some economic
or demographic characteristics of the municipalities. For that purpose, we
consider a generalization of model (\ref{def:alpharAM}) in which
interactions between variables are allowed. The model selection procedure
allowed us to retain only two significant variablesto fit $\alpha _{it}^{r},$
in (\ref{def:alpharAM}): the initial level of employment (\texttt{SIZE}) of
the municipality and its population density (\texttt{DENSITY}). Using an
approximate ANOVA test procedure (see \citealp{MR2206355}), an additive
structure for $\alpha _{it}^{r}$, \textit{i.e} $\alpha _{it}^{r}=\alpha
_{t}^{r}+$\ $g_{1t}^{r}(\mbox{\texttt{SIZE}}_{i})+g_{2t}^{r}(%
\mbox{\texttt{DENSITY}}_{i})$ is strongly rejected for all years $t$ in
favor of a more general model based on bivariate regression functions, 
\begin{equation*}
\alpha _{it}^{r}=\alpha _{t}^{r}+g_{t}^{r}(\mbox{\texttt{SIZE}}_{i},%
\mbox{\texttt{DENSITY}}_{i}),\quad r=1,\ldots ,R-1.
\end{equation*}

As seen in Table~\ref{table_main} in the Supplementary file, a first result
that emerges is that the estimates of the parametric part of (\ref%
{def:alpharAM}), representing the mean effect of the policies for the
subsamples $\{i\ |Y_{it}^{D_{i}}-Y_{it_{0}}\neq 0\},$ i.e. $\alpha _{t}^{r},$
indicate a very short-run (\textit{abrupt but transitory}) and quite low
mean effect of ZRR. Indeed, the estimated value of $\alpha _{t}^{ZRR}$ for
the pre-program years is close to zero and is clearly not significant; then,
it grows and rises up to $2.159$ ({\normalsize \texttt{p-values}}$=0.063$) \
when $t=1999$. Afterwards, it sharply decreases and becomes close to zero
again at the end of the period. It is instead highlighted a \textit{gradual
start, long-term duration }mean effect for the joint 5B-ZRR treatment since  
$\widehat{\alpha }_{t}^{5ZRR\&5B}$ grows overtime, reaching the pick of $%
3.537$ ({\normalsize \texttt{p-values}}$=0.001$) \ when $t=1999$  and then
it slowly decreases over time. Finally, $\widehat{\alpha }_{it}^{5B}$ has a
similar time pattern than $\widehat{\alpha }_{t}^{ZRR\&5B}$ but it is not
significant at standard levels. 

\begin{figure}[H]
\centering               
\subfloat[ZRR]{
  \includegraphics[width=130mm]{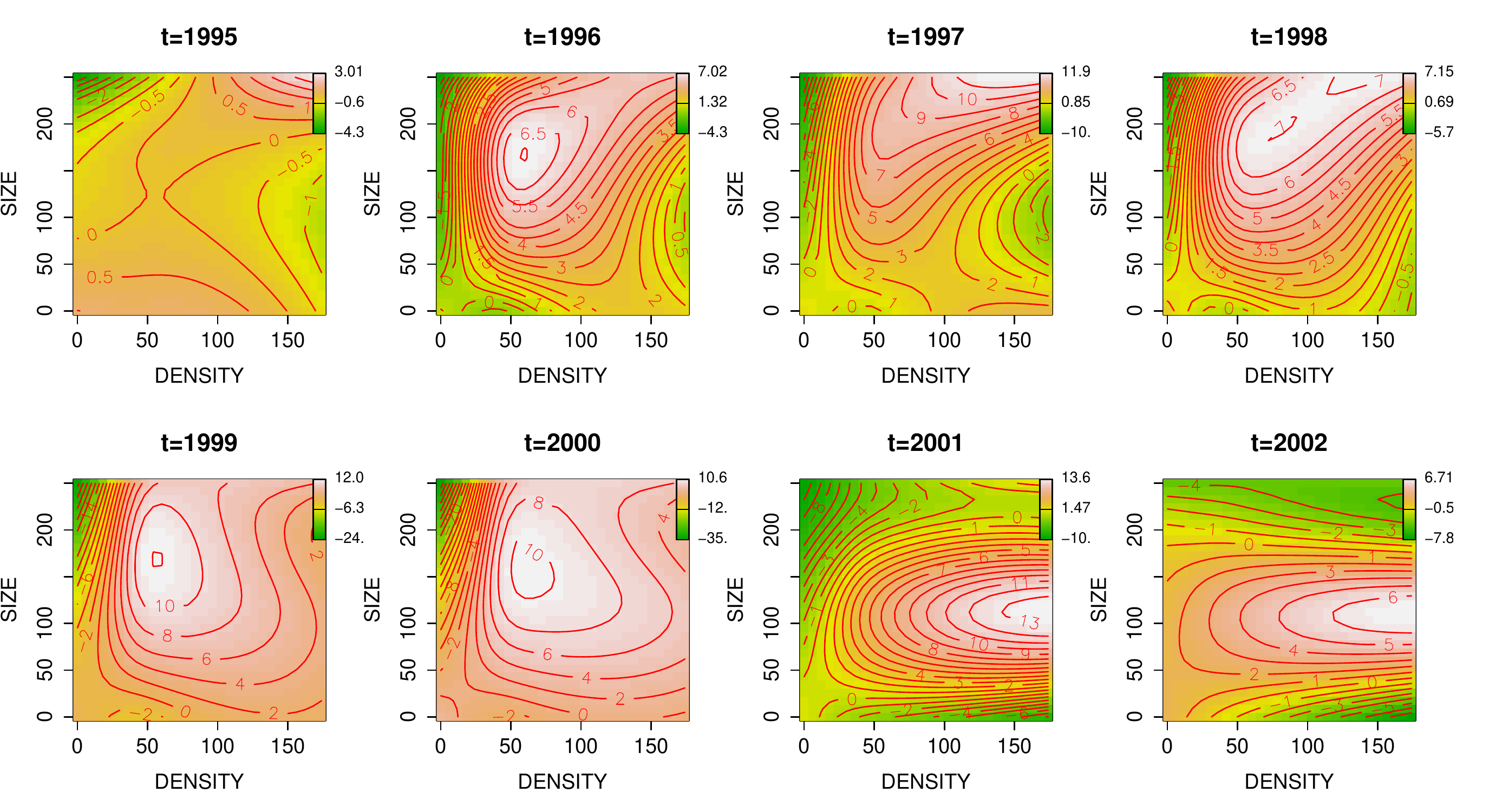}
}
\par
\subfloat[ZRR\&5B]{
  \includegraphics[width=130mm]{CONTOUR_ZRR.pdf}
} \hspace{0mm} 
\subfloat[5B]{
  \includegraphics[width=130mm]{CONTOUR_ZRR.pdf}
}
\caption{Contour plots of $\widehat{g}_{t}^{r}(\mathtt{SIZE},\mathtt{DENSITY}%
).$}
\label{plot_g}
\end{figure}

Next, the examination of the nonparametric part $g_{t}^{r}\left( .\right) $
of (\ref{def:alpharAM}), reveals how the mean impact of the policy $r$
varies as a function of the characteristics in terms of density and size of
each considered statistical unit (see Figure \ref{plot_g}).

\begin{center}
\textbf{Figure \ref{plot_g} about here}
\end{center}

In almost all cases, the smooth functions appear to be highly significant,
using a Bayesian approach to variance estimation \citep{wood2012p}, with
generally quite high effective degrees of freedom, thus indicating rather
complex functions (see \citealp{MR2206355}). 
For all the treatments, we first note that both the magnitude and the shape
of the nonparametric effect vary with time. Looking at ZRR, the estimated
smooth function $\widehat{g}_{t}^{ZRR}(\mathtt{SIZE}_{i}\mathtt{,DENSITY}%
_{i})$ is very flat and close to zero at the beginning and at the end of the
period whereas it becomes clearly nonlinear with a bell-shaped pattern for a
period of a few years after the introduction of the policy. The maximum of
these functions is generally reached for levels of \texttt{DENSITY} slightly
above 50 and for levels of \texttt{SIZE} at about 150, even if the location
of these maxima slightly change over time. For the last two years, the
maximum is reached for slightly smaller and denser municipalities. Note that
in the plots, the domain of \texttt{SIZE }and \texttt{DENSITY} has been
appropriately reduced to focus on municipalities not having a too large
sizes or very high levels of density. The joint nonparametric effect of ZRR
and 5B, $\widehat{g}_{t}^{ZRR\&5B}(\mathtt{SIZE}_{i}\mathtt{,DENSITY}_{i})$
behaves similarly in terms of shape and time pattern but with a stronger
effect for the years 1999 and 2000. Finally, the estimated nonparametric
surface measuring the effect of 5B, $\widehat{g}_{t}^{5B}(\mathtt{SIZE}_{i}%
\mathtt{,DENSITY}_{i})$, is generally quite flat, even if some positive
effects appeared for rather low levels of \texttt{DENSITY} and for $t\geq
1999.$

\subsubsection*{Modeling the probability of no variation}

\label{sec:logit}

Generalized additive models based on binomial regression with logit link
function are fitted to estimate the probability that a variation of the
response between $t$ and $t_{0}$ does not occur, given the treatment status
and the initial conditions. This conditional probability is expressed in (%
\ref{def:gamlogit}).

Again, a backward variable selection procedure has been employed to select
the variables to be introduced in the model. The estimation results are
presented in Table~\ref{table_main} and indicate that the 5B program has a
negative effect on the probability that employment does not vary along time.
The estimated parameter $\widehat{\delta \beta }_{0t}^{5B}$ is always
negative, in a significant way for nearly all instants $t$. Referring to (%
\ref{def:gamlogit}), this means that $\mathbb{P}\left[ \Delta
_{it}^{5B}=0\mid X_{i}\right] -\mathbb{P}\left[ \Delta _{it}^{0}=0\mid X_{i}%
\right] <0$. \ When, looking at ZRR, it can be noted that the estimated
parameter $\widehat{\delta \beta }_{0t}^{ZRR}$ is always positive, but is
not significant in most of the cases, so that $\mathbb{P}\left[ \Delta
_{it}^{ZRR}=0\mid X_{i}\right] -\mathbb{P}\left[ \Delta _{it}^{0}=0\mid X_{i}%
\right] $ is not significantly different from zero. Finally, the estimated
joint policy effect $\widehat{\delta \beta }_{0t}^{ZRR\&5B}$ is always very
close to zero and is never significant.

As far as the mean differential effect, $\mathbb{E}\left[
Y_{it}^{r}-Y_{it}^{0}\mid X_{i}\right] $ in (\ref{def:effectzero}) is
concerned, these results suggest that, for both ZRR and the joint policy
ZRR\&5B, this differential effect is mostly affected by the first part of
the expression, i.e. by $\left( 1-\mathbb{\ P}\left[ \Delta _{it}^{r}=0\mid
X_{i}\right] \right) \times \alpha _{t}^{r}(X_{i})$. Conversely, for 5B,
there is an additional effect arising from the second part of the
expression, since, as noted before, $\mathbb{P}\left[ \Delta
_{it}^{5B}=0\mid X_{i}\right] -\mathbb{P}\left[ \Delta _{it}^{0}=0\mid X_{i}%
\right] <0$.

\subsection{Main results}

\subsubsection*{Counterfactual analysis at an individual level}

We now provide the main results of the estimation and combine
information from the two parts, the continuous and the discrete one, of the mixture model. Notably, our nonparametric approach
allows for non-linear and local effects and thus make it possible to conduct
a temporal counterfactual analysis at an individual level. This relevant
feature is illustrated on a few representative municipalities for which the
evolutions of the potential outcomes are estimated and compared under the
different possible treatments. These municipalities have been chosen with a
clustering partition around medoids procedure (see %
\citealp{KaufmanRousseeuw1990}) with four clusters so that they represent
four different homogeneous groups. Descriptive statistics are given in Table~%
\ref{table_conter}.

\begin{center}
\textbf{Table \ref{table_conter} about here}
\end{center}

Using (\ref{eq:deltar}), (\ref{eq:delta0}), (\ref{def:alpharAM}) and (\ref%
{def:gamlogit}) we can estimate what would have been the evolution of the
expected effect of each municipality under each policy, taking account of
the zero inflation effect. We are also interested in building confidence
intervals. Due to the complexity of our statistical estimations at an
individual level, which are products of predictions obtained with
generalized additive models, the standard delta method cannot be used
easily. We consider instead the more flexible bootstrap approach (see 
\textit{e.g.} \citealp{EfronBook}) to approximate the distribution of the
conditional counterfactual outcome of each selected municipality $i$ having
characteristics $X_{i}$.

We draw $B=1000$ bootstrap samples and for each bootstrap sample $b$, with $%
b=1,\ldots B$, we make the following estimation of the expected
counterfactual evolution~(see~(\ref{def:effectzero})), 
\begin{align*}
\widehat{\mathbb{E}}^{b}\left[ Y_{it}^{r}-Y_{it}^{0}\mid X_{i}\right] &
=\left( 1-\widehat{\mathbb{P}}^{b}\left[ \Delta _{it}^{r}=0\mid X_{i}\right]
\right) \times \widehat{\alpha }_{t}^{r,b}(X_i) \\
& -\left( \widehat{\mathbb{P}}^{b}\left[ \Delta _{it}^{r}=0\mid X_{i}\right]
-\widehat{\mathbb{P}}^{b}\left[ \Delta _{it}^{0}=0\mid X_{i}\right] \right)
\times \widehat{\mu}_{t}^{0,b}(X_{i}),
\end{align*}%
where $\widehat{\mathbb{P}}^{b}\left[ \Delta _{it}^{r}=0\mid X_{i}\right] $
is the estimated probability, with sample $b$, of no employment variation
and $\widehat{\alpha }_{t}^{r,b}(X_i)$ and $\widehat{\mu}_{t}^{0,b}(X_{i})$
are the fitted values. Then, we can deduce, using the percentile method,
bootstrap confidence intervals for the conditional expectation $\mathbb{E}%
\left[ Y_{it}^{r}-Y_{it}^{0}\mid X_{i}\right] $, \textit{i.e} the mean
effect at time $t$ on municipality with characteristics $X_{i}$ of treatment 
$r$ compared to no treatment.

\subsubsection*{Temporal policy effects for the selected municipalities}

Estimated expected counterfactual values as well as bootstrap confidence
intervals are drawn in Figure~\ref{fig:counter} for the four municipalities
under study. The first selected municipality, which is named DSI1, is an
extremely dense and urbanized municipality, with values of \texttt{DENSITY}
and \texttt{URB} greater than the 95th percentile. It is also very rich in
terms of \texttt{INCOME} and big in terms of \texttt{SIZE}, with values of
these variables about the 80th percentile. For this municipality, we
estimate a positive evolution of employment in the absence of any policy. We
can also note that, according to our model, ZRR, 5B and the joint policies
ZRR\&5B would have no significant effect on the evolution of employment for
the considered period. . 

The second municipality, named DS3, is rather dense, urbanized and big, with
values of \texttt{DENSITY}, \texttt{URB and SIZE} about the 75th percentile
of our sample. The value of \texttt{INCOME} is close to the median. We note
that the effect of 5B is quite low -- and is only significant for $t=1997$
and $t=1998$ -- and presents a rather flat evolution over time, whereas both
ZRR and the joint policy ZRR\&5B have a higher impact on employmememt, with
an inverted U pattern. Such an impact increases over the years reaching a
peak for $t=1999$ and then it slowly decreases during the following years.

\begin{center}
\textbf{Figure \ref{fig:counter} about here}
\end{center}

The third municipality, DSI3, is quite close to the median values in terms
of \texttt{DENSITY} and \texttt{SIZE}. For this municipality, all the
policies produce an effect with an inverted U time pattern, even if the
effect of ZRR is significant only for a short period, i.e. over 1998-2001.
Finally, for the last municipality DSI7, which is a small and poor
municipality, there is a clear positive effect of 5B over all the period
(except the last year), with again an inverted U pattern over time. For this
municipality, ZRR has instead no significant effect over the whole period.

These results highlight that ZRR and 5B are likely to produce temporal
effects that vary according to the typology of the municipalities. Indeed,
while the structural funds 5B are effective for very small and rural
municipalities, the fiscal incentives through ZRR produce an effect for
bigger and more dense/urbanized areas. This result is consistent with the
idea that agglomeration externalities \citep{devereux2007firm} and an
adequate size of the local market are essential in order to make such fiscal
incentives effective, while the structural funds, which mainly cover
investments in infrastructure, technology and productive assets, may produce
an effect even for very deprived areas. Finally, the lack of effects for
extremely dense and big municipalities is not surprising because only few
municipalities with such characteristics are treated and the policies under
investigation have not been designed for such a typology of municipality.

Our results can be seen as a refinement of previous studies focusing on
average effects. As far as the French experience is concerned, \cite%
{BehaghelLorenceau2015} did not find any significant average effect of ZRR
at a canton level over the period, even if they underlyine that \textit{%
\textquotedblleft this lack of effect may hide positive impacts on some
specific segments\textquotedblright } (\citealp[p. 9]{BehaghelLorenceau2015}%
).  It can be also noted that beyond the French experience, the literature
generally provides mixed evidence. In some papers a significant effect on
employment (\citealp{Papke1994,ham2011government}) is noted for such
policies whereas some other works indicate that EZ have been ineffective (%
\citealp{bondonio2000enterprise,neumark2010enterprise}), or only provide a
transitory effect \ \citep{Keefe2004}. Moreover, as far as the structural
funds are concerned, \cite{becker2010going} focus attention on the effect of
Objective 1 on regional growth for NUTS2 and NUTS3 regions and find evidence
of temporal effects, with an average effect that takes four years to become
significant and increases afterwards up to the sixth and last available year
after its introduction. Overall, we provide evidence that allowing the
effect of the policy to vary in time and across municpalities can be useful
to show the existence of temporal effects over short periods of time for
some specific municipalities, which otherwise could be missed when looking
at average effects over time. The next subsection will provide further
insights on such an issue.

\subsubsection*{A comparison with standard parametric approaches}

As a final step, we compare the proposed approach with some standard
methods, which are based on parametric models or which do not account for
the mass at zero. This may provide relevant insights because, as stressed
for instance by \cite{lechner2011estimation}, the size of the bias of
misspecified parametric models can be assessed only through a comparison.
The models we consider are listed below:

\begin{itemize}
\item Model 1: Linear continuous response model with homogeneous temporal
effect, $\mu _{t}^{0}(X_{i})=\mu _{t}^{0}+\sum_{j=1}^{p}\beta
_{jt}^{0}X_{ij};$ $\alpha _{it}^{r}=\alpha _{t}^{r}.$

\item Model 2: Linear continuous response model with linear policy
interactions, $\mu _{t}^{0}(X_{i})=\mu _{t}^{0}+\sum_{j=1}^{p}\beta
_{jt}^{0}X_{ij};$ $\alpha _{it}^{r}=\alpha _{t}^{r}+\gamma _{t}^{r}\mathtt{%
SIZE}_{i}+\theta _{t}^{r}\mathtt{DENSITY}_{i}.$

\item Model 3: Linear mixture distribution model with linear policy
interactions, $\mu _{t}^{0}(X_{i})=\mu _{t}^{0}+\sum_{j=1}^{p}\beta
_{jt}^{0}X_{ij};$ $\alpha _{it}^{r}=\alpha _{t}^{r}+\gamma _{t}^{r}\mathtt{%
SIZE}_{i}+\theta _{t}^{r}\mathtt{DENSITY}_{i}$.

\item Model 4: Additive mixture distribution model with nonparametric policy
interactions, $\mu _{t}^{0}(X_{i})=\mu
_{t}^{0}+\sum_{j=1}^{p}g_{jt}^{0}(X_{ij});$ $\alpha _{it}^{r}=\alpha
_{t}^{r}+g_{t}^{r}(\mathtt{SIZE}_{i},\mathtt{DENSITY}_{i})$.
\end{itemize}

The first model is a continuous parametric response model. It is simple
extension of the difference-in-differences estimator that allows for
temporal policy effects and that takes account of linear effects of the
initial conditions. This model is very standard in the policy evaluation
literature. Also, note that in Model 1, \ the policy effect does not change
across municipalities. The second one allows the term $\alpha _{it}^{r}$ to
be a linear function of \texttt{SIZE} and \texttt{DENSITY}. The third one
extends the previous one by handling the zero inflation effect. Finally, the
fourth model is the one we propose in this paper allowing for additive
smooth effects of the initial conditions, nonparametric policy interactions
and handling the zero inflation phenomenon.

\begin{center}
\textbf{Figure \ref{fig:counter2} about here}
\end{center}

As an illustrative example, we focus on municipality DSI3, whose values of 
\texttt{SIZE} and \texttt{DENSITY} are close to the median. The results are
depicted in Figure \ref{fig:counter2}. As far as Model~1 is concerned, we
note that none of the policies is found to provide a significant effect, the
only exception being the joint 5B-ZRR policy at time $t=1999$. This result
appears to be in sharp contrast with the results from the proposed Model 4,
which highlights nonlinear and significant temporal effects for all the
treatments. We can also remark the big difference concerning the estimated
employment under no treatment when comparing the two approaches. Then, when
moving to Model 2/Model 3, some temporal effects appear, even if, by
imposing a parametric policy interaction, $\alpha _{it}^{r}=\alpha
_{t}^{r}+\gamma _{t}^{r}\mathtt{SIZE}_{i}+\theta _{t}^{r}\mathtt{DENSITY}_{i}
$, we obtain very different time patterns of the estimated effects compared
to the ones obtained with the more flexible Model 4. 
Specifically, while models 2 and 3 indicate a monotonically increasing
overtime effect of 5B and an increasing effect of ZRR, with a threshold for
the last years in the sample, Model 4 suggests an inverted U pattern for
both policies. Finally, it is worth comparing Model 2 with Model 3, where
the only difference is the fact of accounting or not for the mass of
observations at zero. When comparing these two models, it can be noted that
the policy effects present similar temporal patterns but handling the zero
inflation feature of the data makes increase the estimated policy effect of
about 15\% -20\%. The same result is obtained when comparing Model 4 with a
similar model that does not account for the mass of observations at zero.We
finally note that allowing for nonlinear effects of the initial conditions
greatly affects the estimates of the policy effects. As an example, when we
estimate a model similar to model 4 by imposing the restriction $\alpha
_{it}^{r}=\alpha _{t}^{r}$, we find evidence of significant temporal
effects, while when we also impose linear effects of the initial conditions,
we do not find any significant effect as in Model 1. 

\section{Concluding remarks}

In this paper, we introduce a semi-parametric approach to estimate the
variation along time and across municipalities of regional treatment effects
in France. Since we face a kind of  zero inflated phenomenon that cannot be
dealt properly with a continuous distribution, we consider a mixture distribution
model that combines a Dirac mass at zero and a continuous response. We rely
on additive models for the continuous response and generalized additive
models for modeling the probabiltiy of a mass at zero, giving more
flexibility than linear models, and we exploit the longitudinal structure of
the data to account for selection bias.

We find that the different policies under investigation are likely to
produce temporal effects that vary according to the typology of the
municipalities. We also documented that using the proposed semi-parametric
model that allows the effect of the policy to vary in time and across
municipalities is crucial to show the existence of temporal effects over
short periods of time for some specific municipalities, which otherwise will
be missed when using standard parametric approaches. We finally provide
evidence that accounting for the mass of observation at zero is important to
avoid a substantial underestimation of the effect of the policies.

This work provides new results about the pattern of temporal treatment
effects and nonlinear interactions, as well as some guidance for future
research. It first suggests, within a flexible semi-parametric regression
framework, a way to deal with an excess of zeros by considering a mixture of 
a continuous and a discrete distribution. This may be relevant for other
policy evaluations when the dependent variable does not vary along time for
a non-negligible fraction of the units. Second, the consideration of a model
in which the effect of the policy is expanded as a nonparametric function of
the covariates provides a richer framework that allows for a finer analysis
and permits to perform a counterfactual estimation at an individual level.
This could be relevant in many cases in which heterogeneous policy effects
are likely to be present or when there is an interest in units having some
peculiar characteristics.

Finally note that the proposed model is flexible and modular enough so that
it can be extended in various directions. In our opinion, an extremely
relevant issue concerns the possible existence of policy effects on
neighboring municipalities, i.e. spatial spillover effects (see e.g. %
\citealp{BehaghelLorenceau2015}). Appendix C in the Supplementary material
indicates that using our local approach, instead of focusing on average
effects, can be crucial to highlight the existence of significant spillover
effects and suggests that further studies may deepen such an issue.

\bibliographystyle{chicago}

\begin{thebibliography}{}

\bibitem[\protect\citeauthoryear{Angrist and Hahn}{Angrist and
  Hahn}{2004}]{angrist2004control}
Angrist, J. and J.~Hahn (2004).
\newblock When to control for covariates? panel asymptotics for estimates of
  treatment effects.
\newblock {\em Review of Economics and statistics\/}~{\em 86\/}(1), 58--72.

\bibitem[\protect\citeauthoryear{Becker, Egger, and Von~Ehrlich}{Becker
  et~al.}{2010}]{becker2010going}
Becker, S.~O., P.~H. Egger, and M.~Von~Ehrlich (2010).
\newblock Going {NUTS}: The effect of {EU} structural funds on regional
  performance.
\newblock {\em Journal of Public Economics\/}~{\em 94\/}(9), 578--590.

\bibitem[\protect\citeauthoryear{Behaghel, Lorenceau, and Quantin}{Behaghel
  et~al.}{2015}]{BehaghelLorenceau2015}
Behaghel, L., A.~Lorenceau, and S.~Quantin ({2015}, {MAY}).
\newblock {Replacing churches and mason lodges? Tax exemptions and rural
  development}.
\newblock {\em Journal of Public Economics\/}~{\em {125}}, {1--15}.

\bibitem[\protect\citeauthoryear{Bondonio and Engberg}{Bondonio and
  Engberg}{2000}]{bondonio2000enterprise}
Bondonio, D. and J.~Engberg (2000).
\newblock Enterprise zones and local employment: evidence from the states'
  programs.
\newblock {\em Regional Science and Urban Economics\/}~{\em 30\/}(5), 519--549.

\bibitem[\protect\citeauthoryear{Brown, Earle, and Telegdy}{Brown
  et~al.}{2006}]{brown2006productivity}
Brown, J.~D., J.~S. Earle, and A.~Telegdy (2006).
\newblock The productivity effects of privatization: Longitudinal estimates
  from {H}ungary, {R}omania, {R}ussia, and {U}kraine.
\newblock {\em Journal of political economy\/}~{\em 114\/}(1), 61--99.

\bibitem[\protect\citeauthoryear{Devereux, Griffith, and Simpson}{Devereux
  et~al.}{2007}]{devereux2007firm}
Devereux, M.~P., R.~Griffith, and H.~Simpson (2007).
\newblock Firm location decisions, regional grants and agglomeration
  externalities.
\newblock {\em Journal of Public Economics\/}~{\em 91\/}(3-4), 413--435.

\bibitem[\protect\citeauthoryear{Efron and Tibshirani}{Efron and
  Tibshirani}{1993}]{EfronBook}
Efron, B. and R.~J. Tibshirani (1993).
\newblock {\em An introduction to the bootstrap}, Volume~57 of {\em Monographs
  on Statistics and Applied Probability}.
\newblock Chapman and Hall, New York.

\bibitem[\protect\citeauthoryear{Friedlander and Robins}{Friedlander and
  Robins}{1995}]{friedlander1995evaluating}
Friedlander, D. and P.~K. Robins (1995).
\newblock Evaluating program evaluations: New evidence on commonly used
  nonexperimental methods.
\newblock {\em The American Economic Review\/}, 923--937.

\bibitem[\protect\citeauthoryear{Frolich}{Frolich}{2004}]{Frolich2004}
Frolich, M. ({2004}, {APR}).
\newblock {Programme evaluation with multiple treatments}.
\newblock {\em Journal of Economic Surveys\/}~{\em {18}\/}({2}), {181--224}.

\bibitem[\protect\citeauthoryear{Gobillon, Magnac, and Selod}{Gobillon
  et~al.}{2012}]{GobillonMagnac2012}
Gobillon, L., T.~Magnac, and H.~Selod ({2012}, {OCT}).
\newblock {Do unemployed workers benefit from enterprise zones? The French
  experience}.
\newblock {\em Journal of Public Economics\/}~{\em {96}\/}({9-10}), {881--892}.

\bibitem[\protect\citeauthoryear{Ham, Swenson, {\.I}mrohoro{\u{g}}lu, and
  Song}{Ham et~al.}{2011}]{ham2011government}
Ham, J.~C., C.~Swenson, A.~{\.I}mrohoro{\u{g}}lu, and H.~Song (2011).
\newblock Government programs can improve local labor markets: Evidence from
  state enterprise zones, federal empowerment zones and federal enterprise
  community.
\newblock {\em Journal of Public Economics\/}~{\em 95\/}(7), 779--797.

\bibitem[\protect\citeauthoryear{Hastie and Tibshirani}{Hastie and
  Tibshirani}{1990}]{MR1082147}
Hastie, T.~J. and R.~J. Tibshirani (1990).
\newblock {\em Generalized additive models}, Volume~43 of {\em Monographs on
  Statistics and Applied Probability}.
\newblock Chapman and Hall, Ltd., London.

\bibitem[\protect\citeauthoryear{Heckman and Hotz}{Heckman and
  Hotz}{1989}]{HeckmanHotz}
Heckman, J. and V.~Hotz (1989).
\newblock Choosing among alternative nonexperimental methods for estimating the
  impact of social programs: the case of manpower training.
\newblock {\em J. Amer. Statist. Assoc.\/}~{\em 84}, 862--874.

\bibitem[\protect\citeauthoryear{Heckman, Lalonde, and Smith}{Heckman
  et~al.}{1999}]{HLS99}
Heckman, J., R.~Lalonde, and J.~Smith (1999).
\newblock The economics and econometrics of active labor market programs.
\newblock In O.~Ashenfelter and D.~Card (Eds.), {\em Handbook of Labor
  Economics}, Volume~3, Chapter~31, pp.\  1865--2097. Elsevier.

\bibitem[\protect\citeauthoryear{Hsiao, Ching, and Wan}{Hsiao
  et~al.}{2011}]{HsiaoChing2011}
Hsiao, C., H.~S. Ching, and S.~Wan (2011).
\newblock A panel data approach for program evaluation-measuring the impact of
  political and economic interaction of {H}ong-{K}ong with {M}ainland {C}hina.
\newblock {\em Journal of Applied Econometrics\/}~{\em 27}, 705--740.

\bibitem[\protect\citeauthoryear{Hubert and Vandervieren}{Hubert and
  Vandervieren}{2008}]{hubert2008adjusted}
Hubert, M. and E.~Vandervieren (2008).
\newblock An adjusted boxplot for skewed distributions.
\newblock {\em Computational statistics \& data analysis\/}~{\em 52\/}(12),
  5186--5201.

\bibitem[\protect\citeauthoryear{Imai and Van~Dyk}{Imai and
  Van~Dyk}{2004}]{imai2004causal}
Imai, K. and D.~A. Van~Dyk (2004).
\newblock Causal inference with general treatment regimes: Generalizing the
  propensity score.
\newblock {\em Journal of the American Statistical Association\/}~{\em
  99\/}(467), 854--866.

\bibitem[\protect\citeauthoryear{Imbens and Wooldridge}{Imbens and
  Wooldridge}{2009}]{ImbensWooldridge2009}
Imbens, G.~W. and J.~M. Wooldridge ({2009}, {MAR}).
\newblock {Recent Developments in the Econometrics of Program Evaluation}.
\newblock {\em Journal of Economic Literature\/}~{\em {47}\/}({1}), {5--86}.

\bibitem[\protect\citeauthoryear{Kaufman and Rousseeuw}{Kaufman and
  Rousseeuw}{1990}]{KaufmanRousseeuw1990}
Kaufman, L. and P.~J. Rousseeuw (1990).
\newblock {\em Finding groups in data}.
\newblock Wiley Series in Probability and Mathematical Statistics: Applied
  Probability and Statistics. John Wiley \& Sons, Inc., New York.
\newblock An introduction to cluster analysis, A Wiley-Interscience
  Publication.

\bibitem[\protect\citeauthoryear{Lechner}{Lechner}{2011a}]{lechner2011estimation}
Lechner, M. (2011a).
\newblock The estimation of causal effects by difference-in-difference methods.
\newblock {\em Foundations and Trends{\textregistered} in Econometrics\/}~{\em
  4\/}(3), 165--224.

\bibitem[\protect\citeauthoryear{Lechner}{Lechner}{2011b}]{Lechner2011}
Lechner, M. (2011b).
\newblock The relation of different concepts of causality used in time series
  and microeconometrics.
\newblock {\em Econometric Reviews\/}~{\em 30}, 109--127.

\bibitem[\protect\citeauthoryear{Lechner}{Lechner}{2015}]{Lechner2015}
Lechner, M. (2015).
\newblock Treatment effects and panel data.
\newblock In B.~Baltagi (Ed.), {\em The Oxford Handbook of Panel Data}. Oxford
  University Press.

\bibitem[\protect\citeauthoryear{Lee}{Lee}{2005}]{lee2005micro}
Lee, M.-J. (2005).
\newblock {\em Micro-econometrics for policy, program, and treatment effects}.
\newblock Oxford University Press on Demand.

\bibitem[\protect\citeauthoryear{Li and Ruppert}{Li and
  Ruppert}{2008}]{li2008asymptotics}
Li, Y. and D.~Ruppert (2008).
\newblock On the asymptotics of penalised splines.
\newblock {\em Biometrika\/}, 415--436.

\bibitem[\protect\citeauthoryear{Marra and Wood}{Marra and
  Wood}{2011}]{MarraWood2011}
Marra, G. and S.~N. Wood (2011).
\newblock Practical variable selection for generalized additive models.
\newblock {\em Comput. Statist. Data Anal.\/}~{\em 55\/}(7), 2372--2387.

\bibitem[\protect\citeauthoryear{McLachlan and Peel}{McLachlan and
  Peel}{2000}]{Mclachlan2000}
McLachlan, G.~J. and D.~Peel (2000).
\newblock {\em Finite Mixture Models}.
\newblock New York: Wiley Series in Probability and Statistics.

\bibitem[\protect\citeauthoryear{Mohl and Hagen}{Mohl and
  Hagen}{2010}]{mohl2010eu}
Mohl, P. and T.~Hagen (2010).
\newblock Do {EU} structural funds promote regional growth? new evidence from
  various panel data approaches.
\newblock {\em Regional Science and Urban Economics\/}~{\em 40\/}(5), 353--365.

\bibitem[\protect\citeauthoryear{Neumark and Kolko}{Neumark and
  Kolko}{2010}]{neumark2010enterprise}
Neumark, D. and J.~Kolko (2010).
\newblock Do enterprise zones create jobs? evidence from californian enterprise
  zone program.
\newblock {\em Journal of Urban Economics\/}~{\em 68\/}(1), 1--19.

\bibitem[\protect\citeauthoryear{O'Keefe}{O'Keefe}{2004}]{Keefe2004}
O'Keefe, S. (2004).
\newblock Job creation in {C}alifornians enterprise zones: A comparison using a
  propensity score matching model.
\newblock {\em Journal of Urban Economics\/}~{\em 55}, 131--150.

\bibitem[\protect\citeauthoryear{Papke}{Papke}{1994}]{Papke1994}
Papke, L. (1994).
\newblock Tax policy and urban development. evidence from the indiana
  enterprise zone program.
\newblock {\em Journal of Public Economics\/}~{\em 54}, 37--49.

\bibitem[\protect\citeauthoryear{Peters and Fisher}{Peters and
  Fisher}{2004}]{PetersFisher2004}
Peters, A. and P.~Fisher (2004).
\newblock The failures of economic development incentives.
\newblock {\em Journal of the American Planning Association\/}~{\em 70},
  27--37.

\bibitem[\protect\citeauthoryear{Rosenbaum and Rubin}{Rosenbaum and
  Rubin}{1983}]{RosenbaumRubin1983}
Rosenbaum, P. and D.~Rubin (1983).
\newblock The central role of the propensity score in observational studies for
  causal effects.
\newblock {\em Biometrika\/}~{\em 70}, 41--55.

\bibitem[\protect\citeauthoryear{Rosenbaum}{Rosenbaum}{2002}]{rosenbaum2002observational}
Rosenbaum, P.~R. (2002).
\newblock Observational studies.
\newblock In {\em Observational Studies}, pp.\  1--17. Springer.

\bibitem[\protect\citeauthoryear{Rousseeuw, Ruts, and Tukey}{Rousseeuw
  et~al.}{1999}]{rousseeuw1999bagplot}
Rousseeuw, P.~J., I.~Ruts, and J.~W. Tukey (1999).
\newblock The bagplot: a bivariate boxplot.
\newblock {\em The American Statistician\/}~{\em 53\/}(4), 382--387.

\bibitem[\protect\citeauthoryear{Ruppert, Wand, and Carroll}{Ruppert
  et~al.}{2003}]{ruppert03}
Ruppert, D., M.~P. Wand, and R.~J. Carroll (2003).
\newblock Semiparametric regression.
\newblock In {\em Cambridge series in statistical and probabilistic
  mathematics}. Cambridge university press.

\bibitem[\protect\citeauthoryear{Sheather}{Sheather}{2004}]{sheather2004density}
Sheather, S.~J. (2004).
\newblock Density estimation.
\newblock {\em Statistical Science\/}~{\em 19\/}(4), 588--597.

\bibitem[\protect\citeauthoryear{Silverman}{Silverman}{1986}]{Silverman86}
Silverman, B.~W. (1986).
\newblock {\em Density estimation for statistics and data analysis}.
\newblock Monographs on Statistics and Applied Probability. Chapman \& Hall,
  London.

\bibitem[\protect\citeauthoryear{Wood}{Wood}{2004}]{MR2090902}
Wood, S.~N. (2004).
\newblock Stable and efficient multiple smoothing parameter estimation for
  generalized additive models.
\newblock {\em J. Amer. Statist. Assoc.\/}~{\em 99\/}(467), 673--686.

\bibitem[\protect\citeauthoryear{Wood}{Wood}{2006}]{Wood2006low}
Wood, S.~N. (2006).
\newblock Low-rank scale-invariant tensor product smooths for generalized
  additive mixed models.
\newblock {\em Biometrics\/}~{\em 62\/}(4), 1025--1036.

\bibitem[\protect\citeauthoryear{Wood}{Wood}{2008}]{wood2008fast}
Wood, S.~N. (2008).
\newblock Fast stable direct fitting and smoothness selection for generalized
  additive models.
\newblock {\em Journal of the Royal Statistical Society: Series B (Statistical
  Methodology)\/}~{\em 70\/}(3), 495--518.

\bibitem[\protect\citeauthoryear{Wood}{Wood}{2012}]{wood2012p}
Wood, S.~N. (2012).
\newblock On p-values for smooth components of an extended generalized additive
  model.
\newblock {\em Biometrika\/}~{\em 100\/}(1), 221--228.

\bibitem[\protect\citeauthoryear{Wood}{Wood}{2017}]{MR2206355}
Wood, S.~N. (2017).
\newblock {\em Generalized additive models: an introduction with R, second
  edition}.
\newblock Texts in Statistical Science Series. Chapman \& Hall/CRC, Boca Raton,
  FL.

\bibitem[\protect\citeauthoryear{Wood, Pya, and S{\"a}fken}{Wood
  et~al.}{2016}]{Wood2016}
Wood, S.~N., N.~Pya, and B.~S{\"a}fken (2016).
\newblock Smoothing parameter and model selection for general smooth models.
\newblock {\em J. Amer. Statist. Assoc.\/}~{\em 111}, 1548--1563.

\bibitem[\protect\citeauthoryear{Wooldridge}{Wooldridge}{2005}]{Wooldridge2005}
Wooldridge, J.~M. (2005).
\newblock Fixed-effects and related estimators for correlated random
  coefficient and treatment-effect panel data models.
\newblock {\em The Review of Economics and Statistics\/}~{\em 87}, 395--390.

\end{thebibliography}

\begin{thebibliography}{Hubert and Vandervieren, 2008}
\bibitem[Becker et~al., 2010]{becker2010going} Becker, S.~O., Egger, P.~H.,
and Von~Ehrlich, M. (2010). \newblock Going {NUTS}: The effect of {EU}
structural funds on regional performance. 
\newblock {\em Journal of Public
Economics}, 94(9):578--590.

\bibitem[Behaghel et~al., 2015]{BehaghelLorenceau2015} Behaghel, L.,
Lorenceau, A., and Quantin, S. ({2015}). 
\newblock {Replacing churches and mason lodges? Tax exemptions and rural
  development}. \newblock {\em Journal of Public Economics}, {125}:{1--15}.

\bibitem[Brown et~al., 2006]{brown2006productivity} Brown, J.~D., Earle,
J.~S., and Telegdy, A. (2006). \newblock The productivity effects of
privatization: Longitudinal estimates from {H}ungary, {R}omania, {R}ussia,
and {U}kraine. \newblock {\em Journal of political economy}, 114(1):61--99.

\bibitem[Friedlander and Robins, 1995]{friedlander1995evaluating} %
Friedlander, D. and Robins, P.~K. (1995). \newblock Evaluating program
evaluations: New evidence on commonly used nonexperimental methods. %
\newblock {\em The American Economic Review}, 85:923--937.

\bibitem[Gobillon et~al., 2012]{GobillonMagnac2012} Gobillon, L., Magnac,
T., and Selod, H. ({2012}). 
\newblock {Do unemployed workers benefit from enterprise zones? The French
  experience}. \newblock {\em Journal of Public Economics}, {96}({9-10}):{%
881--892}.

\bibitem[Heckman and Hotz, 1989]{HeckmanHotz} Heckman, J. and Hotz, V.
(1989). \newblock Choosing among alternative nonexperimental methods for
estimating the impact of social programs: the case of manpower training. %
\newblock {\em J. Amer. Statist. Assoc.}, 84:862--874.

\bibitem[Heckman et~al., 1999]{HLS99} Heckman, J., Lalonde, R., and Smith,
J. (1999). \newblock The economics and econometrics of active labor market
programs. \newblock In Ashenfelter, O. and Card, D., editors, \emph{Handbook
of Labor Economics}, volume~3, chapter~31, pages 1865--2097. Elsevier.

\bibitem[Hubert and Vandervieren, 2008]{hubert2008adjusted} Hubert, M. and
Vandervieren, E. (2008). \newblock An adjusted boxplot for skewed
distributions. \newblock {\em Computational statistics \& data analysis},
52(12):5186--5201.

\bibitem[Lee, 2005]{lee2005micro} Lee, M.-J. (2005). 
\newblock {\em
Micro-econometrics for policy, program, and treatment effects}. \newblock %
Oxford University Press on Demand.

\bibitem[Rosenbaum, 2002]{rosenbaum2002observational} Rosenbaum, P.~R.
(2002). \newblock Observational studies. \newblock In \emph{Observational
Studies}, pages 1--17. Springer.

\bibitem[Rousseeuw et~al., 1999]{rousseeuw1999bagplot} Rousseeuw, P.~J.,
Ruts, I., and Tukey, J.~W. (1999). \newblock The bagplot: a bivariate
boxplot. \newblock {\em The American Statistician}, 53(4):382--387.

\bibitem[Wood, 2006]{Wood2006low} Wood, S.~N. (2006). \newblock Low-rank
scale-invariant tensor product smooths for generalized additive mixed
models. \newblock {\em Biometrics}, 62(4):1025--1036.

\bibitem[Wood et~al., 2016]{Wood2016} Wood, S.~N., Pya, N., and S{\"a}fken,
B. (2016). \newblock Smoothing parameter and model selection for general
smooth models. \newblock {\em J. Amer. Statist. Assoc.}, 111:1548--1563.

\bibitem[Wooldridge, 2005]{Wooldridge2005} Wooldridge, J.~M. (2005). %
\newblock Fixed-effects and related estimators for correlated random
coefficient and treatment-effect panel data models. 
\newblock {\em The
Review of Economics and Statistics}, 87:395--390.
\end{thebibliography}

\newpage

\begin{figure}[H]
\begin{center}
\begin{tabular}{lllll}
&  & $D$ & $\longrightarrow $ & $Y_{t}^r-Y_{t_{0}}$ \\ 
& $\mathbf{\nearrow }$ & $\uparrow $ & $\nearrow $ & $\uparrow $ \\ 
$X_{\bar{Y}\cap D}$ &  & $X_{Y\cap D}$ &  & $X_{Y\cap \bar{D}}$%
\end{tabular}%
\end{center}
\par
\label{causality}
\caption{The expected causal relation between $Y_{t}^{r}-Y_{t_{0}}$, $X$ and 
$D$}
\label{fig:causality}
\end{figure}

\begin{center}
\begin{figure}[H]
\centering               
\subfloat[DSI1]{
  \includegraphics[width=85mm]{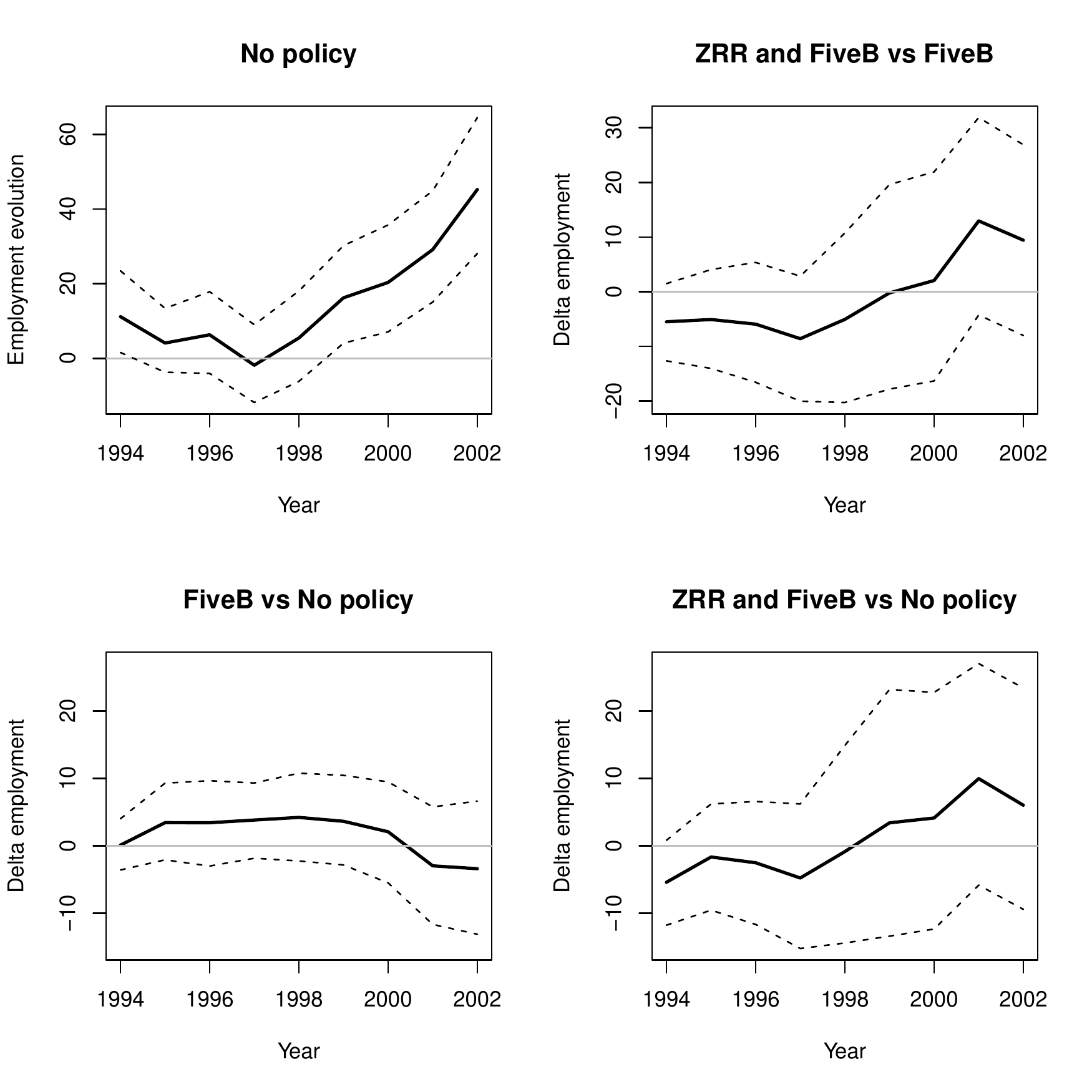}
} \subfloat[DS3]{
  \includegraphics[width=85mm]{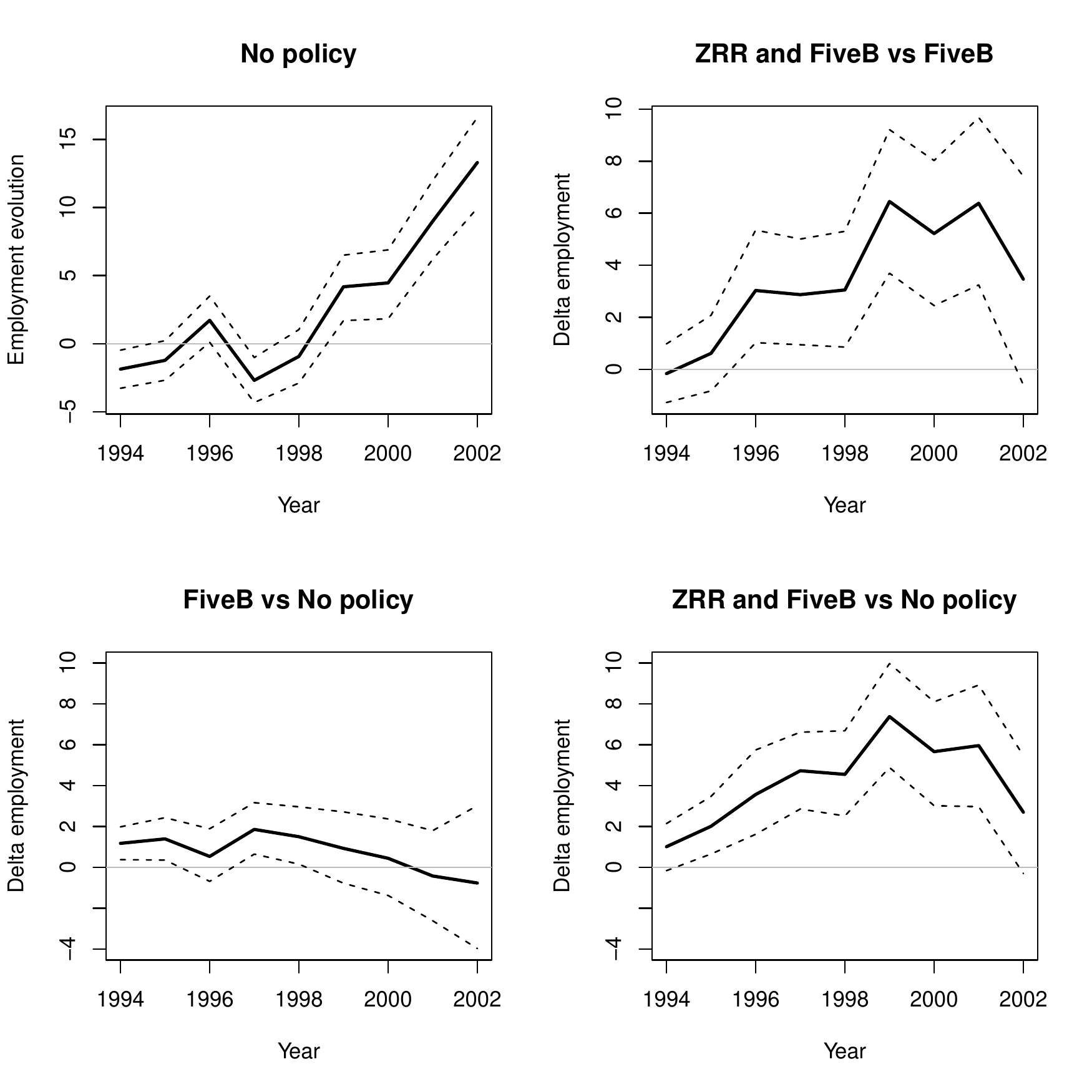}
} \hspace{0mm} 
\subfloat[DSI3]{
  \includegraphics[width=85mm]{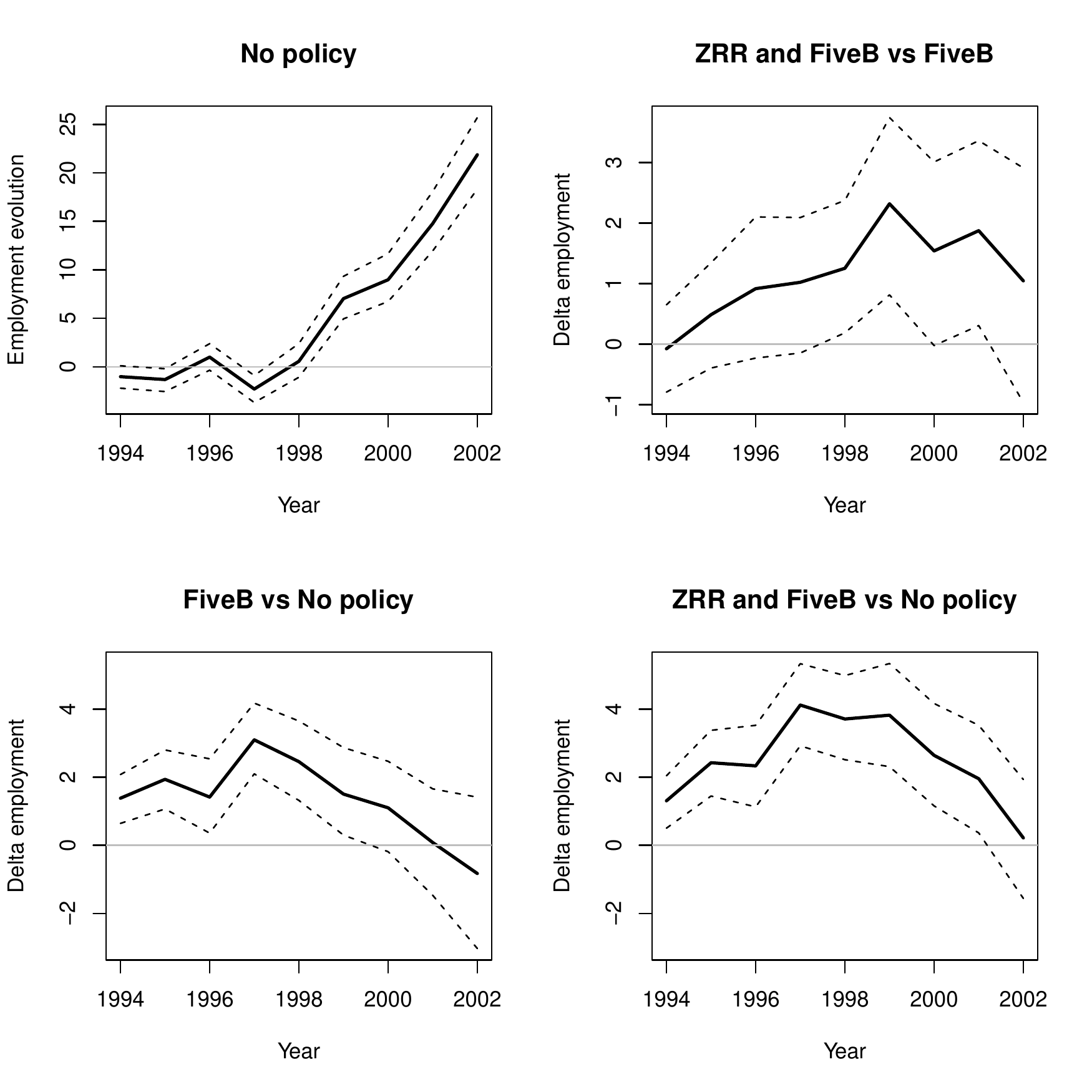}
} \subfloat[DSI7]{
  \includegraphics[width=85mm]{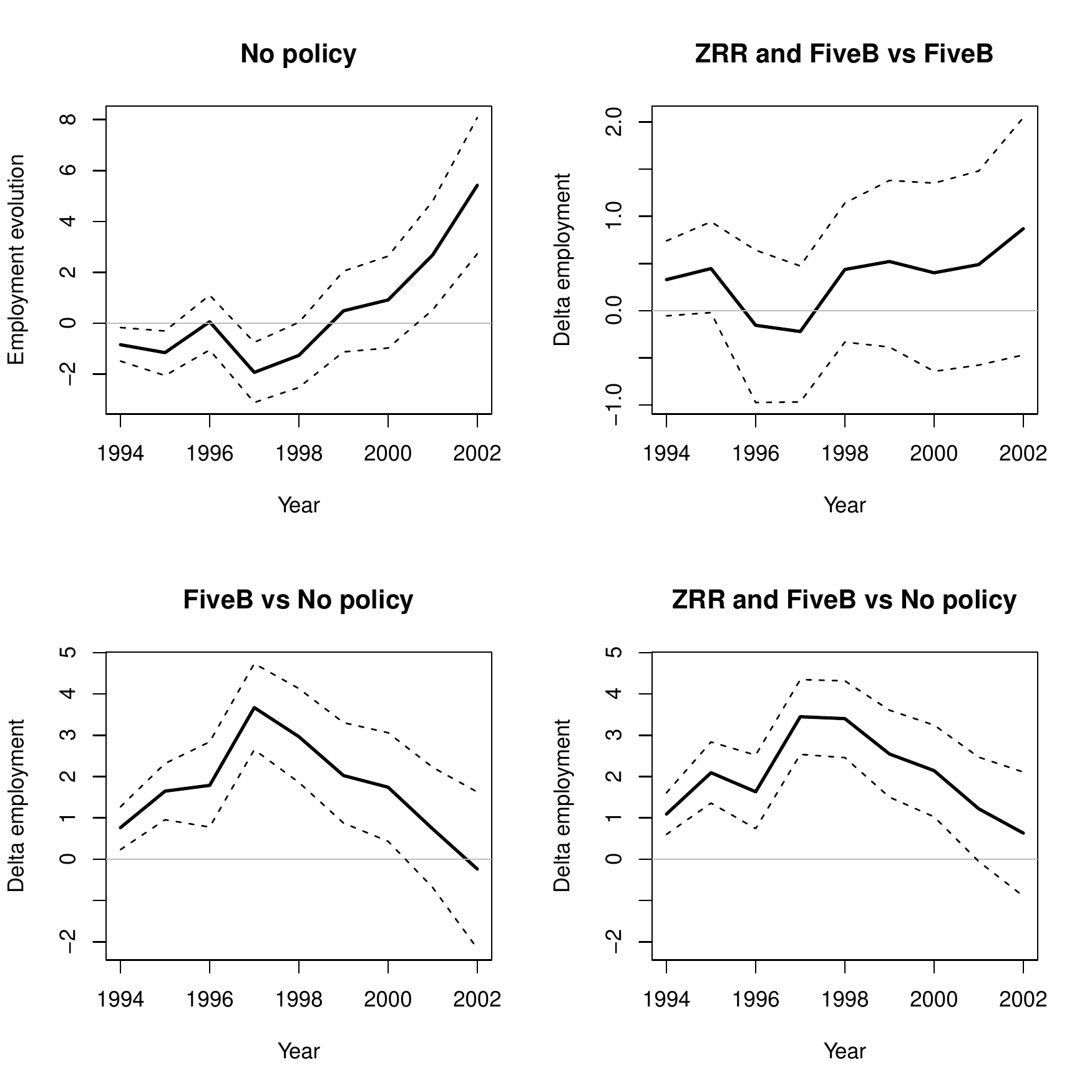}
}
\caption{{\protect\small Counterfactual estimation of the evolution of the
employment level for the selected municipalities. The first plot (top left)
represents the estimated evolution of employment when no funds are given to
the municipality. The others plots represent the difference of evolution
between the joint policies ZRR and Five B compared to only Five B (top
right), between Five B and no policy (bottom left) and between the joint
policies ZRR and Five B compared to no policy at all. Mean values are drawn
in plain line and 95\% bootstrap confidence intervals in dotted line. }}
\label{fig:counter}
\end{figure}

\bigskip

\begin{figure}[H]
\centering               
\subfloat[Model 1]{
  \includegraphics[width=85mm]{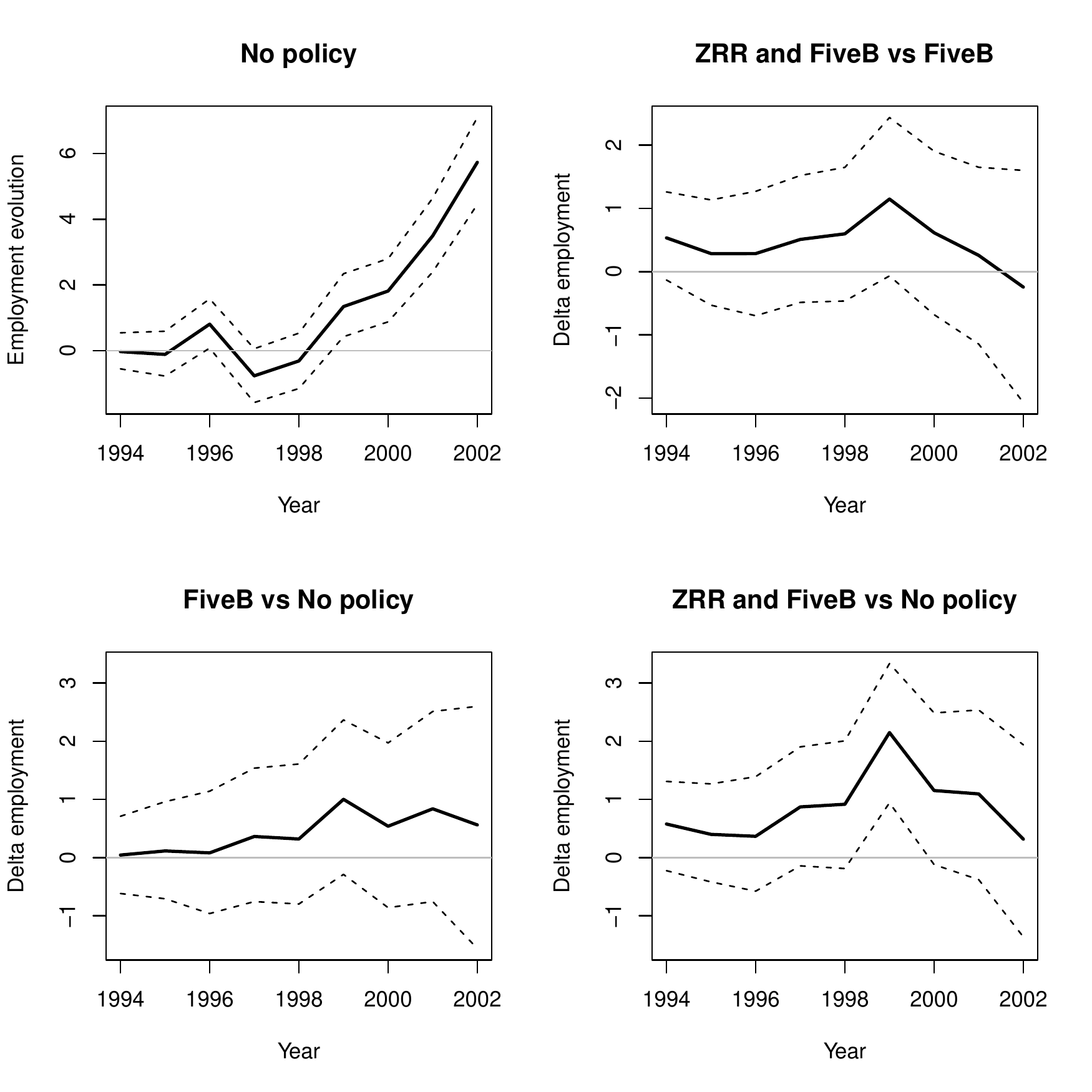}
} \subfloat[Model 2]{
  \includegraphics[width=85mm]{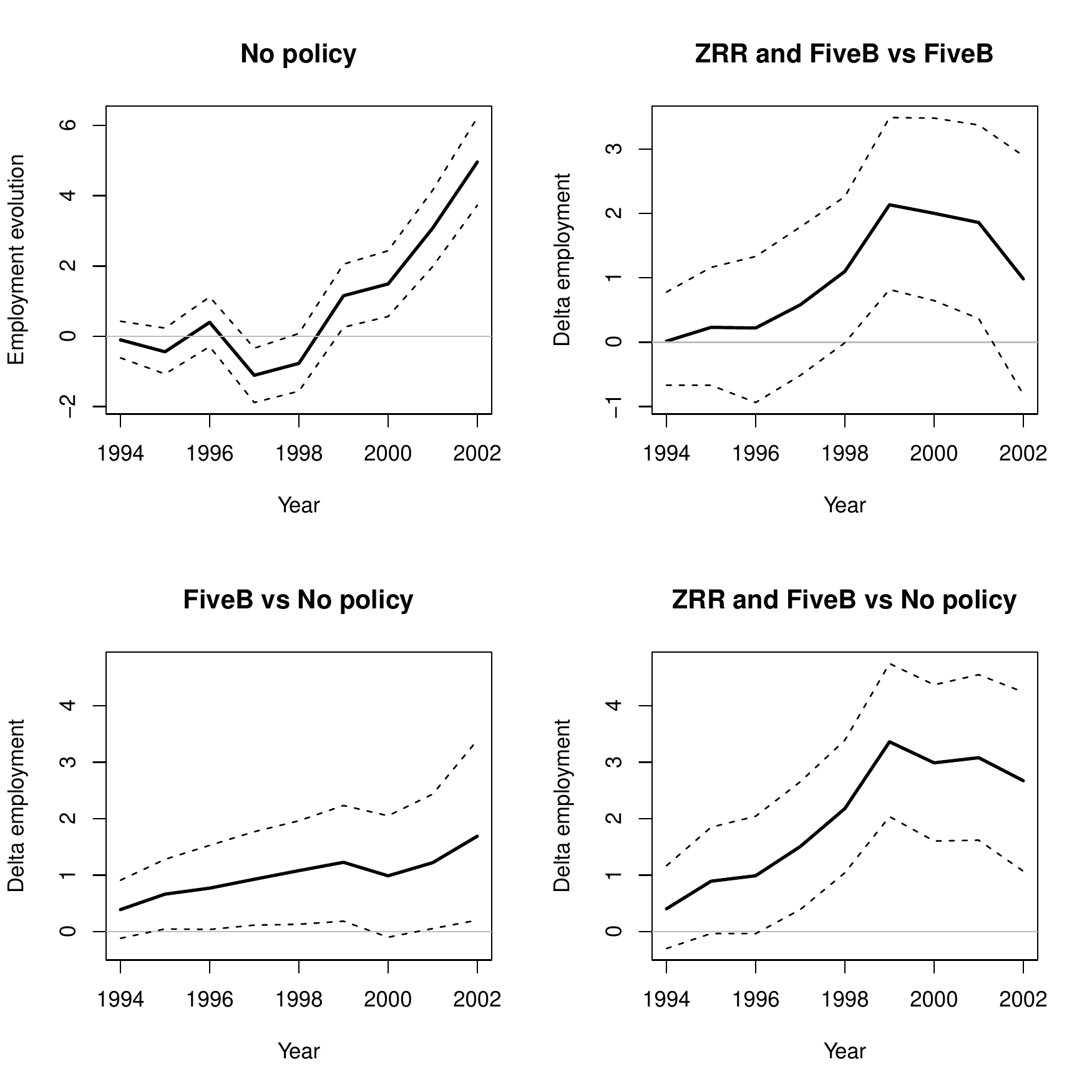}
} \hspace{0mm} 
\subfloat[Model 3]{
  \includegraphics[width=85mm]{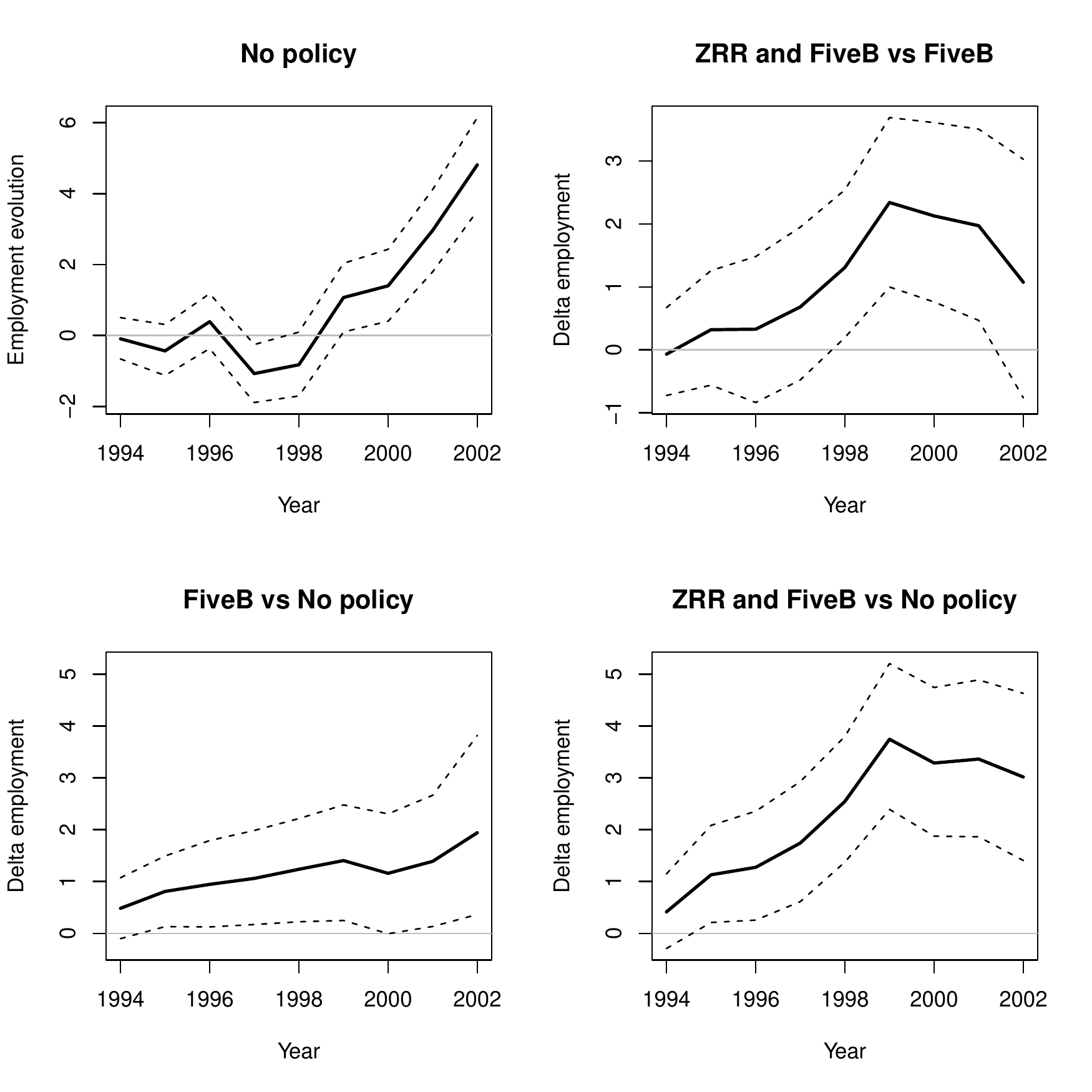}
} \subfloat[Model 4]{
  \includegraphics[width=85mm]{CounterDSI3.pdf}
}
\caption{{\protect\small Counterfactual estimation of the evolution of the
employment level for the selected municipality DSI3. Model 1: Linear
continuous response model, no interactions, }$\protect\alpha _{it}^{r}=%
\protect\alpha _{t}^{r}${\protect\small . Model 2: Linear continuous
response model with linear policy interactions, }$\protect\alpha _{it}^{r}=%
\protect\alpha _{t}^{r}+\protect\gamma _{t}^{r}SIZE_{i}+\protect\theta %
_{t}^{r}DENSITY_{i}${\protect\small . Model 3: Linear mixture distribution
model with linear policy interactions, }$\protect\alpha _{it}^{r}=\protect%
\alpha _{t}^{r}+\protect\gamma _{t}^{r}SIZE_{i}+\protect\theta %
_{t}^{r}DENSITY_{i}${\protect\small . Model 4: Additive mixture distribution
model with nonparametric policy interactions, }$\protect\alpha _{it}^{r}=%
\protect\alpha _{t}^{r}+g_{t}^{r}(SIZE_{i},DENSITY_{i})${\protect\small .
Mean values are drawn in plain line and 95\% bootstrap confidence intervals
in dotted line. } }
\label{fig:counter2}
\end{figure}

\bigskip



\bigskip 

\bigskip 

\begin{table}[H]
\begin{center}
\begin{tabular}{ccccccccc}
Municip. & DENSITY & SIZE & INCOME & OLD & FACT & BTS & AGRIH & URB \\ \hline
DSI1 & 218.85 & 105 & 5772 & 0.11 & 0.19 & 0.016 & 0.08 & 0.23 \\ 
DS3 & 61.26 & 48 & 4324 & 0.30 & 0.06 & 0.037 & 0.19 & 0.032 \\ 
DSI3 & 41.87 & 25 & 6300 & 0.20 & 0.13 & 0.038 & 0.03 & 0.028 \\ 
DSI7 & 22.74 & 10 & 3724 & 0.14 & 0.16 & 0.007 & 0.22 & 0.015%
\end{tabular}%
\end{center}
\caption{Descriptive statistics of the municipalities selected for
counterfactual analysis.}
\label{table_conter}
\end{table}
\end{center}

\bigskip \newpage

\bigskip

\newpage \appendix

\noindent {\huge \textbf{Supplementary Material} }

\bigskip

\section{Data, variables and sample}

\setcounter{table}{0} \renewcommand{\thetable}{A\arabic{table}}

\setcounter{figure}{0} \renewcommand{\thefigure}{A\arabic{figure}}

{\normalsize After the merge of the different sheets provided by the INSEE
containing information on local employment, the demographic structure,
education and land use, we get a data set containing 36000 municipalities,
that is the 98,5\% of the French municipalities. While the paper focuses
specifically on rural development policies, it is worth recalling that a
relevant fraction of the municipalities received structural funds (1994-99)
not specifically devoted to rural development. These are the Objective 1 and
the Objective 2 funds. Objective 1 has the explicit aim of fostering per
capita GDP growth in regions that are lagging behind the EU average -
defined as those areas with a per capita GDP of less than 75 per cent of the
EU average - and of promoting aggregate growth in the EU. Objective 2 covers
regions struggling with structural difficulties and aims to reduce the gap
in socio-economic development by financing productive investment in
infrastructures, local development initiatives and business activities.
Table~\ref{table_A} describes the distribution of the municipalities
according to the ZRR and the structural funds schemes (1994-99).}

{\normalsize Among the 646 municipalities under Objective 1, 350 are located
in Corsica. All the Corsica's municipalities available in our dataset are
under the Objective 1. Among them, 268 were also under ZRR scheme. The
remaining 296 municipalities under Objective 1 are located in the region
Nord-Pas de Calais and were not under ZRR. Given the small number of
municipalities under the Objective 1 and their specific characteristics, we
decided to remove them from the analysis. This simplifies greatly the
framework of the analysis without losing a relevant amount of information,
getting a dataset containing 35354 municipalities. }

{\normalsize As far as Objective 2 is concerned, we initially estimated the
proposed model by including a treatment variable defined as $D_{i}\in
\left\{ 0,EU2,5B,ZRR,ZRR\&EU2,ZRR\&5B\right\} $, which also accounts for the
Objective 2, $EU2$ ($ZRR\&EU2)$ indicating that the municipality \textit{i}
receive incentives only from Objective 2 (from both Objective 2 and ZRR).
However, the estimated parameters $\widehat{\alpha }_{t}^{EU2}$\ \ and $%
\widehat{\alpha }_{t}^{ZRR\&EU2}$ ($\widehat{\delta \beta }_{0t}^{EU2}$ and $%
\widehat{\delta \beta }_{0t}^{ZRR\&EU2}$) were always very close to zero and
never significant with \texttt{p-values} very far from standard significance
levels. This result along with the fact that the interest of this paper is
on rural development, motivated the use of the treatment variable defined in
Section~IV where the Objective 2 municipalities are considered as if they
had not received any treatment. The use of such a variable simplifies the
analysis and the presentation of the results without losing relevant
information, also provided that the four parameters of interest $\alpha
_{t}^{5B},$ $\alpha _{t}^{ZRR\&5B}$ and $\alpha _{t}^{ZRR}$ ($\delta \beta
_{0t}^{5B},$ $\delta \beta _{0t}^{ZRR\&5B}$ and $\delta \beta _{0t}^{ZRR}$)
are fundamentally not affected by such a choice. }

{\normalsize The dependent variable $Y_{it}$ ($i$ indicating the
municipality; and $t$ the time $t=1993,...,2003$) measures the number of
employees and has been calculated from the SIRENE data sheet covering
manufacture, trade and services, while the initial full set of regressors
(measured at time $t=1990$) is composed of the following 16 variables: }

{\normalsize \bigskip }

{\normalsize \noindent \textbf{Initial outcome} }

{\normalsize $\mathtt{SIZE}_{i}\equiv Y_{it_{0}}$ is the initial outcome,
i.e the level of employment at $t_{0},$ with $t_{0}$ equals to $1993$; }

{\normalsize \bigskip }

{\normalsize \noindent \textbf{Socio-economic and demographic variables} }

{\normalsize $\mathtt{DENSITY}_{i}\equiv \left( \text{\textit{population}}%
\right) _{i}/\left( \text{\textit{surface in terms of \ km}}^{2}\right)
_{i}; $ }

{\normalsize $\mathtt{OLD}_{i}\equiv \left( \text{\textit{population over 65}%
}\right) _{i}/\left( \text{\textit{total population}}\right) _{i};$ }

{\normalsize $\mathtt{INC}_{i}\equiv \left( \text{\textit{net taxable income}%
}\right) _{i}/\left( \text{\textit{total population}}\right) _{i};$ }

{\normalsize $\mathtt{FACT}_{i}\equiv \left( \text{\textit{number\ of} 
\textit{factory\ workers}}\right) _{i}/\left( \text{\textit{total\ population%
}}\right) _{i}$\textit{;} }

{\normalsize $\mathtt{EXE}_{i}\equiv \left( \text{\textit{number\ of} 
\textit{executive\ workers}}\right) _{i}/\left( \text{\textit{total\
population}}\right) _{i}$\textit{;} }

{\normalsize $\mathtt{FARM}_{i}\equiv \left( \text{\textit{number\ of} 
\textit{\ farmers}}\right) _{i}/\left( \text{\textit{total\ population}}%
\right) _{i}$\textit{;} }

{\normalsize $\mathtt{UNIV}_{i}$ $\equiv \frac{\left( \text{\textit{number\
of\ people\ with\ a\ master\ level\ degree called \textquotedblleft Ma\^{\i}%
trise universitaire\textquotedblright }}\right) _{i}}{\left( \text{\textit{%
total\ population}}\right) _{i}}\ \ $\textit{;} }

{\normalsize $\mathtt{BTS}_{i}\equiv \frac{\left( \text{\textit{number\ of\
people\ with\ a\ \ technical} \textit{degree called\ \textquotedblleft
Brevet de Technicien Sup\'{e}rieur\textquotedblright }}\right) _{i}}{\left( 
\text{\textit{total\ population}}\right) _{i}}$\textit{;} }

{\normalsize $\mathtt{NOEDU}_{i}$ $\equiv \left( \text{\textit{number\ of\
people\ without\ a} \textit{degree}}\right) _{i}/\left( \text{\textit{total\
population}}\right) _{i}$\textit{;} }

{\normalsize \bigskip }

{\normalsize \noindent \textbf{Land use} }

{\normalsize $\mathtt{AGRI}_{i}\equiv $ $\left( \text{\textit{farmland} 
\textit{surface}}\right) _{i}/\left( \text{\textit{total\ surface}}\right)
_{i}$; }

{\normalsize $\mathtt{CULT}_{i}\equiv $ $\left( \text{\textit{cultivated land%
} \textit{surface}}\right) _{i}/\left( \text{\textit{total\ surface}}\right)
_{i}$; }

{\normalsize $\mathtt{URB}_{i}\equiv $ $\left( \text{\textit{urban surface}}%
\right) _{i}/\left( \text{\textit{total\ surface}}\right) _{i}$; }

{\normalsize $\mathtt{IND}_{i}\equiv $ $\left( \text{\textit{industrial
surface}}\right) _{i}/\left( \text{\textit{total\ surface}}\right) _{i}$; }

{\normalsize $\mathtt{ARA}_{i}\equiv $ $\left( \text{\textit{arable surface}}%
\right) _{i}/\left( \text{\textit{total\ surface}}\right) _{i}$; }

{\normalsize $\mathtt{GRA}_{i}\equiv $ $\left( \text{\textit{grassland
surface}}\right) _{i}/\left( \text{\textit{total\ surface}}\right) _{i}$; }

{\normalsize \bigskip }

{\normalsize The socio-economic and demographic variables come from standard
INSEE sources while the variables measuring land use have been obtained from
the \textquotedblleft Corine Land Cover\textquotedblright\ base (providing
remote sensing images which have been merged with the French map at a
municipality level). }

{\normalsize \bigskip }

{\normalsize The retained models, those results are presented in Section~4,
have been obtained using a backward selection procedure starting from the
above set of potential explanatory variables. Backward selection provided
almost the same results as the double penalty approach proposed by Marra and
Wood (2011), those detailed results are available upon request. \ More
precisely, we selected the variables equation-by-equation for $%
t=1994,...,2002,$ by setting the threshold level for the \texttt{p-values}
to $0.01$ and in the end, to use the same explanatory variables \ for all $t$%
, we choosed the variables that were 1\% significant at least for one time
period, $t$. According to the notation used in Section 3, these variables
are noted as $X_{Y\cap D}$ and $X_{Y\cap \bar{D}}.$ }

{\normalsize For the estimation of the conditional probability of a null
employment variation along time which is expressed in eq. (\ref{def:gamlogit}%
), we retained the following variables:} {\small 
\begin{equation*}
X_{i}^{\{\text{\textit{logit}}\}}=\left(
SIZE_{i},DENSITY_{i},UNIV,INC,FACT_{i},EXE,FARM,BTS_{i},NOEDU,ARA_{i},URB_{i},IND_{i},GRA\right) ,
\end{equation*}%
} {\normalsize while for the continuous part of the model referring to the
subsample $\{i\ |Y_{it}^{D_{i}}-Y_{it_{0}}\neq 0\}$, \ the variables that we
selected are:} 
\begin{equation*}
X_{i}^{\{\text{\textit{continuous}}\}}=\left(
SIZE_{i},DENSITY_{i},OLD_{i},INC,FACT_{i},BTS_{i},CULT_{i},AGRI_{i},ARA_{i},URB_{i},IND_{i}\right) .
\end{equation*}

{\normalsize Tables~\ref{table:statdesc} and \ref{table_desc2} provide
simple descriptive statistics. }

{\normalsize Also note that according to Heckman and Hotz (1989, pg. 865),
selection bias may also arise from the presence of variables that may
influence selection into the program even if they do not affect directly the
outcome and introducing these variables into the regression solves this
additional source of selection bias. Using the notation employed in Section
3, these variables are noted as $X_{\bar{Y}\cap D}.$ We determine these
variables by exploiting recent advances in generalized additive models
permitting the estimation of multinomial logistic regression~\citep{Wood2016}%
. This allows a flexible estimation of a generalized propensity score $%
\mathbb{P}\left[ D_{i}\mid X_{i}\right] $ as a function of additive smooth
components. Again we used the backward selection and finally we added 3 more
variables that appeared to affect selection into the programs and that were
not selected directly from the outcome equation. These variables are $%
FARM_{i}$, $NOEDU_{i}$ and $GRA_{i}.$ \ However, adding these variables does
not produce relevant changes to the estimates of the effects and detailed
results are available upon request. }

{\normalsize Finally, let broadly recall the trimming procedure we used to
determine the sample for the estimation. We dropped outlier observations
which have been identified using a variety of methods such as the visual
inspection of the distribution via kernel density estimation, standard
boxplot, adjusted boxplot for skewed distributions \citep{hubert2008adjusted}%
, bivariate inspection and bivariate boxplot \citep{rousseeuw1999bagplot}.
The variables we collected generally present an asymmetric distribution and
in some cases are characterized by an extremely long right tail. This is the
case of $SIZE_{i}$ (skewness=151) and $DENSITY_{i}$ (skewness=15.69), which
have a crucial role in the model with interactions. For these two variables
we ended as follows. For $SIZE_{i}$, we keep municipalities for which $%
SIZE_{i}<500$, $500$ representing the $92th$ percentile while for $%
DENSITY_{i}$ we select municipalities having $DENSITY_{i}<1000$, 1000 being
about the $97th$ percentile. In both cases, the range of the variable has
been greatly reduced, from $1128000$ to $499$ in the first case and from $%
21940$ to $999$ in the second one. After the cleaning, the sample used for
the estimation contains $25593$ municipalities. For such a sample, we
globally do not observe problems in terms of lack of overlap. This feature
makes the average treatment effect relevant for policy purposes. }

\section{Identification hypotheses and placebo tests}

{\normalsize In order to identify the causal effect, a common practice is to
assume the following hypothesis holds (see e.g. %
\citealp{ImbensWooldridge2009}), }%
\begin{equation}
Y_{it}^{r}\perp \!\!\!\perp D_{i}\ |\ X_{i},\ U_{it}\qquad \forall r\in
\{0,1,\ldots ,R-1\}.  \label{CIA}
\end{equation}%
{\normalsize This general condition means that there exist both observable
variables (}$X_{i}${\normalsize ) and unobservable variables (}$U_{it}$%
{\normalsize ) that are related to the potential outcomes (}$Y_{it}^{r}$%
{\normalsize ) and to the treatment status (}$D_{i}${\normalsize ), such
that given these variables, }$Y_{it}^{r}${\normalsize \ and }$D_{i}$%
{\normalsize \ are independent. This general formulation encompasses the
most widely used specifications in the literature. An important particular
case of the above condition is (\ref{CIA-s}).}

{\normalsize Since selection bias may not be completely eliminated only
after controlling for the observables }$X_{i}${\normalsize , it is also
important to note that a before-after approach may help to address the issue
of selection on unobservables. \ We thus consider (\ref{CIA_B}), which is
more general than (\ref{CIA-s}) and holds for example when the unobservables 
}$U_{it}${\normalsize \ may be described as follows, }%
\begin{equation}
U_{it}=\phi _{1i}+v_{it}  \label{def=twoway}
\end{equation}%
{\normalsize where }$\phi _{1i}${\normalsize \ is a random (individual) time
invariant effect, that may be correlated to the treatment variable }$D_{i}$%
{\normalsize , and }$v_{it}${\normalsize \ is a white noise. }
{\normalsize An alternative specification for the the unobservables is the
so called \textit{random growth model} \citep{HeckmanHotz,Wooldridge2005}, which assumes the following specification for }$%
U_{it}:${\normalsize \ 
\begin{equation}
U_{it}=\phi _{1i}+\phi _{2i}t+v_{it}
\end{equation}%
allowing individual parameters $(\phi _{1i},\phi _{2i})$ to be correlated
with the treatment indicator variable $D_{i}.$To estimate the model, we
adopt the same tranformation as in \cite{HeckmanHotz}, that is $\left[
Y_{it}^{r}-Y_{it_{0}}^{0}-(t-t_{0})\left(
Y_{it_{0}}^{0}-Y_{it_{0}-1}^{0}\right) \right] $ and the underlying
conditional independence assumption on a transformed equation can be written
as }

{\normalsize 
\begin{equation}
\left[ Y_{it}^{r}-Y_{it_{0}}^{0}-(t-t_{0})\left(
Y_{it_{0}}^{0}-Y_{it_{0}-1}^{0}\right) \right] \perp \!\!\!\perp D_{i}\ |\
X_{i},\qquad \forall r\in \{0,1,\ldots ,R-1\}.
\end{equation}%
}

{\normalsize As underlined by \cite{HLS99}, when different methods produce
different inference would suggest that selection bias is important and that
some of the adopted estimators are likely to be misspecified. In order to
detect misspecified models, we implement both `pre-program' and `post
program' tests along the lines depicted by \cite{HeckmanHotz} and
implemented empirically in some previous papers (see e.g. %
\citealp{brown2006productivity,friedlander1995evaluating}). These tests are
based on the idea that a valid estimator would correctly adjust for
differences in pre-program (resp. post-program) outcomes between future
(resp. past) participants and non-participants, otherwise the estimator is
rejected. }

{\normalsize These placebo tests are performed here looking at the effect of
ZRR, because the availability of some years prior the introduction of the
ZRR incentives, occurred in September 1996, allows us to conduct
`pre-program' tests, while for the program 5B, introduced in 1994, there is
not enough statistical information before its introduction. More precisely,
we focus attention on the continuous part on the model, and precisely we
focus on the mean temporal effect of ZRR. We consequently fit a model for $%
\alpha _{it}^{r}$ as in (\ref{def:alpharAM}), assuming that $\alpha
_{it}^{r}=\alpha _{t}^{r}$, for $t_{k}\leq t\leq t_{m}$ and $r\in \{1,\ldots
,R-1\}$.}

{\normalsize The `pre-program' test is generally implemented by setting $t<k$
and by testing the significance of the treatment effect $\alpha _{t}^{r}$.
If $\alpha _{t}^{r}$ is significantly different from $0$ then the underlying
model fails to pass the test$.$ However, even if the logic is compelling, if
a shock or an anticipation effect close to the time of the treatment affects
only one group but not the other, the results from such a test are
potentially misleading. This problem has also been summarized under the
heading \textit{\textquotedblleft fallacy of alignment\textquotedblright }\ %
\citep{HLS99}. In our case, treated firms could (shortly) postpone hiring in
order to obtain the public incentives, so that using quite longer lags can
be useful in order to obtain an effective test and avoiding to overestimate
the treatment effect \citep{brown2006productivity,friedlander1995evaluating}%
. }

{\normalsize Accordingly, we first use all the available information in the
data and use the most distant data before the introduction of the policy to
set $t_{0}$ and propose, for the \textit{before-after} specification, three
tests by setting ($t_{0}=1993$, $t=1994$), ($t_{0}=1993$, $t=1995$) and ($%
t_{0}=1993$, $t=1996$), respectively. Next we set $t_{0}=1994$. This allows
both to make the \textit{before-after} and the \textit{random growth}
estimators directly comparable and to verify the robustness of the previous
tests to a change in the starting point $t_{0}$. }

{\normalsize A post program test has an identical structure to the
pre-program test except that for such a test $t>k$, when neither group
receives the treatment. As for the pre-program test, we alternatively set $%
t_{0}$ to $1993 $ and $1994$, whereas for $t$ we use the last two years in
the sample, that is 2001 and 2002. The interpretation of this kind of test
could be however more problematic than that of the pre-program test since it
could be that a policy has a permanent or a long-term impact on the outcome.
However, the fact that some previous studies pointed out that various EZ
have only a short run impact on employment makes the post-program test of a
certain empirical relevance here. Moreover, even if it cannot be excluded 
\`{a} priori that a rural policy produces an effect only for some few years,
it is difficult to imagine a situation in which its effect become negative
after some years from its adoption. So a negative and significant estimate
of $\alpha _{t}$ for $t>k$ would suggest that the model is misspecified. }

{\normalsize The results from such tests (Table~\ref{table_placebo}) provide
interesting insights which are summarized below. A first relevant result is
that, when analyzing the \textit{before-after }model\textit{,} setting $%
t_{0} $ alternatively to 1993 and 1994 has no effect on the results of the
tests. Secondly, it seems ex-post that the results of the\textit{\
before-after} specification which does not include the initial conditions
are quite unsatisfactory, specifically looking at the post-program tests
since the effect of the policy decreases overtime becoming not only negative
but also statistically significant at the end of the period for $t=2001$ and 
$t=2002$, with \texttt{p-values} very close to zero. Such a negative and
decreasing overtime estimates for the post treatment periods could indicate
that the assumptions underlying the identification of the causal effect are
still too restrictive to obtain a credible result. This could arise because
i) the treated municipalities are expected to have a different (i.e. lower)
time trend than non treated ones even in absence of the policy; \ ii) some
observable factors can be related to the policy placement (also affecting
the outcome variable), those omission from the model causes the so called 
\textit{overt bias}, to adopt the terminology from \cite{lee2005micro} and 
\cite{rosenbaum2002observational}. A third relevant result is that adding
(nonparametrically) the initial conditions greatly improves the results of
the tests (this specification passes both pre and post program tests) and
provides much more credible results. Moreover, non reported results indicate
that using an additive model instead of a linear specification improves
greatly the alignment. }

{\normalsize A central issue concerns the comparison of the \textit{%
before-after} with the \textit{random growth}. If the initial conditions are
not included the \textit{random growth}, similarly to the \textit{%
before-after}, does not pass the post-program tests and provide negative and
decreasing overtime estimates of the treatment effect with with \texttt{%
p-values }below standard levels. When the initial conditions are included,
the results are as follows. While the \textit{before-after} clearly passes
the tests with estimates close to zero and not significant (\texttt{p-values}
are equal to 0.771 and \ 0.847{\small )}, \textit{the random growth} still
provides estimates of $\alpha _{t}$ for the post-program period which are
highly negative (-3.681 and -4.787) and show a decreasing trend overtime
with associated \texttt{p-values} equal to 0.300 and 0.232, which are much
lower than those obtained with the \textit{before-after} specification.
Looking at the estimates for all available $t$ may provide further insights.
The \textit{random growth} provides estimates of the effect of ZRR that are
negative for all $t,$ are relatively high in magnitude and are increasing in
absolute value with $t$. }

{\normalsize These tests suggest the use of a \textit{before-after}
specification added with the initial conditions and allowing for
nonparametric effects of such initial conditions. For such a model, a very
good alignment is obtained pre and post treatment. We do not intend to claim
that we have found the `true' model but a purpose of this paper has been to
reduce the risk of misspecification by relying on semi-parametric modeling
and variable selection and by discarding specifications that fail to provide
a good alignment. }

\section{Extension: spatial spillovers}

\label{sec:spatialspillover}

{\normalsize The proposed model is flexible and modular enough so that it
can be extended in various directions. As an illustrative example, we
address the relevant issue of the possible existence of policy effects on
neighboring municipalities, i.e. spatial spillover effects (see e.g. %
\citealp{BehaghelLorenceau2015}). To save space, the analysis is restricted
to the continuous part of the model. One standard way to deal with this
issue consists in introducing, in the model, explanatory variables
accounting for the absence or the presence of the policies in the
neighboring municipalities. Ex ante, for both ZRR and 5B, the spillovers may
be either positive arising directly through a higher labor demand and/or
indirectly from agglomeration economies or negative if some substitution
effects occur. In practice, the identification of spillovers is an intricate
empirical matter, requiring the definition of the neighborhood and the
choice of an adequate channel of transmission. We focus here on purely
geographic spillovers and adopt a very restrictive notion of neighborhood by
considering the spillovers arising from the municipalities sharing a common
border. Among the 25593 municipalities under study, 10523 municipalities
have all their neighboring municipalities that do not receive any funds,
2496 municipalities have all their neighboring municipalities that are under
5B but not under ZRR while for 239 municipalities, the entire neighborhood
is under ZRR but not under 5B. There is also a group of 7888 municipalities
that have some neighboring municipalities under 5B and some other
neighboring municipalities which are under ZRR. Finally, there is a group of
4447 municipalities with all the neighboring municipalities under both 5B
and ZRR.}

{\normalsize With this classification in mind, we build a new categorical
variable, denoted by WD}$_{i}\in \left\{
0,5\_ALL,Z\_ALL,5\&Z\_SOME,5\&Z\_ALL\right\} ${\normalsize , with modalities
corresponding to the above mentioned categories and the corresponding
parameters are noted }$\omega _{t}^{5\_ALL},\omega _{t}^{Z\_ALL},\omega
_{t}^{5\&Z\_SOME}${\normalsize \ and }$\omega _{t}^{5\&Z\_ALL}${\normalsize %
. These parameters capture the spillover effects by measuring the mean
differential effect, over the whole sample, with respect to the reference
category which is chosen to be }$0${\normalsize , i.e. the category of
municipalities having neighboring municipalities that do not receive any
funds. The new variable WD}$_{i}${\normalsize \ is then added as an
additional explanatory variable in the regression functions given in (\ref%
{def:amy0}) and (\ref{def:alpharAM}). The estimation results indicate no
significant spillover effects, meaning that both ZRR and 5B produced an
effect that remains spatially localized. Geographic spillovers are never
statistically significant with p-values being always very far from standard
significance levels. Note finally that the absence of significant spillover
effects still holds when considering many alternative definitions of WD
based on different considerations about geographic proximity (detailed
results are available upon request). This result is consistent with a recent
literature on regional policy evaluation suggesting that policy spillovers
do not occur or at best, they are modest in magnitude (see e.g. %
\citealp{becker2010going,BehaghelLorenceau2015,GobillonMagnac2012}).}

{\normalsize Interestingly, it appears that if we consider a more flexible
model that allows nonparametric interactions effects we get a different
picture. In particular, some interactive spillovers appear now highly
significant. Note also that after a model selection procedure, the same
variables that have been employed in Section \ref{sec:Results} are retained
in the model, that is SIZE and DENSITY, to interact with WD. We also get
again that an additive structure is rejected in favor of a bivariate smooth
function. This result provides additional empirical support to the
importance of considering flexible models in order to let the data a chance
to speak.}

{\normalsize We finally provide a brief comment to the results presented in
Figure \ref{plot_spillovers}.\footnote{{\normalsize As in previous figures,
the domain of the continuous variables has been appropriately reduced to the
regions where the effects are significant. To save space we focus only on $%
t=1999$.}} For WD}$_{i}\in \left\{ 5\_ALL,5\&Z\_ALL\right\} ${\normalsize ,
we find evidence of significant interactive spillover effects. A relevant
result is that, for both modalities, spillovers are very low or even
negative for low levels of both SIZE and DENSITY, while they become positive
and reach their maximum level for municipalities characterized by high
levels of both variables.}

{\normalsize 
\begin{figure}[]
{\normalsize \centering\includegraphics[width=160mm]{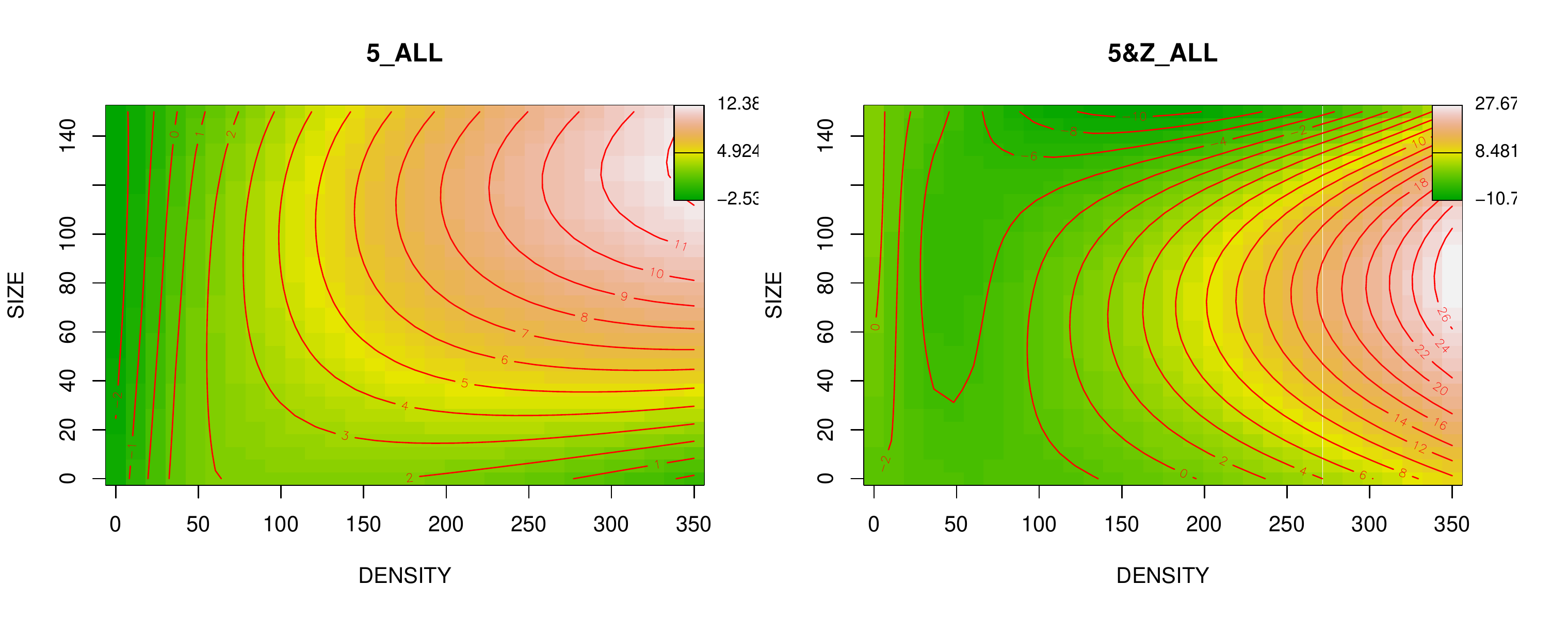} 
}
\caption{Spillover effects. Contour plots.}
\label{plot_spillovers}
\end{figure}
}

{\normalsize \newpage }

\begin{center}
{\normalsize 
\begin{table}[tbp]
{\normalsize 
\begin{tabular}{|l|lll|}
\hline
Structural Funds/ZRR & 0 & 1 & total \\ \hline
0 & 10831 & 401 & 11232 \\ 
1 & 378 & 268 & 646 \\ 
2 & 6815 & 590 & 7405 \\ 
5B & 6641 & 10076 & 16717 \\ 
total & 24665 & 11335 & 36000 \\ \hline
\end{tabular}
\  }
\caption{Distribution of the municipalities according to ZRR and Structural
Funds schemes}
\label{table_A}
\end{table}
}

{\normalsize 
\begin{table}[tbp]
{\normalsize 
\rotatebox{90}{
\begin{tabular}{||lcccccccccc||}
\hline\hline
& \multicolumn{2}{c}{\textbf{FULL}} & \multicolumn{8}{c||}{\textbf{ZONING}}
\\ 
& \multicolumn{2}{c}{\textbf{SAMPLE}} & \multicolumn{8}{c||}{\textbf{SCHEME}}
\\ \hline
& \multicolumn{2}{c}{} & \multicolumn{2}{c}{\textit{0}} & \multicolumn{2}{c}{\textit{5B}} & \multicolumn{2}{c}{\textit{ZRR\&5B}} & \multicolumn{2}{c||}{\textit{ZRR}} \\ \hline
& \multicolumn{2}{c}{} & \multicolumn{2}{c}{} & \multicolumn{2}{c}{} & 
\multicolumn{2}{c}{} & \multicolumn{2}{c||}{} \\ 
n & \multicolumn{2}{c}{\small 25593} & \multicolumn{2}{c}{\small 12580} & 
\multicolumn{2}{c}{\small 5277} & \multicolumn{2}{c}{\small 7014} & 
\multicolumn{2}{c||}{\small 722} \\ 
\% & \multicolumn{2}{c}{} & \multicolumn{2}{c}{\small 0.491} & 
\multicolumn{2}{c}{\small 0.206} & \multicolumn{2}{c}{\small 0.274} & 
\multicolumn{2}{c||}{\small 0.028} \\ \hline
&  &  &  &  &  &  &  &  &  &  \\ 
& mean & s.d. & mean & s.d. & mean & s.d. & mean & s.d. & mean & s.d. \\ 
\hline
SIZE & {\small 56.37} & {\small 92.281} & {\small 68.201} & {\small 101.316}
& {\small 57.22} & {\small 93.453} & {\small 36.665} & {\small 70.317} & 
{\small 35.558} & {\small 67.252} \\ 
DENSITY & {\small 56.02} & {\small 73.590} & {\small 78.084} & {\small 91.599} & {\small 49.603} & {\small 49.364} & {\small 24.215} & {\small 27.592} & 
{\small 27.300} & {\small 33.782} \\ 
OLD & {\small 0.174} & {\small 0.064} & {\small 0.143} & {\small 0.052} & 
{\small 0.179} & {\small 0.056} & {\small 0.222} & {\small 0.060} & {\small 0.190} & {\small 0.057} \\ 
INC & {\small 4838} & {\small 1160.315} & {\small 5302.966} & {\small 1166}
& {\small 4577.307} & {\small 951.730} & {\small 4207.766} & {\small 917.3673} & {\small 4773.478} & {\small 1044.728} \\ 
FACT & {\small 0.141} & {\small 0.053} & {\small 0.152} & {\small 0.050} & 
{\small 0.142} & {\small 0.052} & {\small 0.119} & {\small 0.054} & {\small 0.140} & {\small 0.058} \\ 
BTS & {\small 0.028} & {\small 0.022} & {\small 0.031} & {\small 0.021} & 
{\small 0.027} & {\small 0.020} & {\small 0.026} & {\small 0.022} & {\small 0.024} & {\small 0.022} \\ 
CULT & {\small 0.029} & {\small 0.119} & {\small 0.039} & {\small 0.140} & 
{\small 0.0314} & {\small 0.120} & {\small 0.012} & {\small 0.069} & {\small 0.013} & {\small 0.067} \\ 
AGRI & {\small 0.204} & {\small 0.181} & {\small 0.180} & {\small 0.180} & 
{\small 0.264} & {\small 0.188} & {\small 0.206} & {\small 0.169} & {\small 0.176} & {\small 0.144} \\ 
ARA & {\small 0.325} & {\small 0.306} & {\small 0.421} & {\small 0.319} & 
{\small 0.216} & {\small 0.252} & {\small 0.232} & {\small 0.263} & {\small 0.342} & {\small 0.283} \\ 
URB & {\small 0.022} & {\small 0.039} & {\small 0.032} & {\small 0.046} & 
{\small 0.018} & {\small 0.033} & {\small 0.007} & {\small 0.019} & {\small 0.009} & {\small 0.214} \\ 
IND & {\small 0.001} & {\small 0.005} & {\small 0.001} & {\small 0.007} & 
{\small 0.0008} & {\small 0.005} & {\small 0.0003} & {\small 0.003} & 
{\small 0.0004} & {\small 0.003} \\ \hline\hline
\end{tabular}
}  }
\caption{Descriptive statistics}
\label{table:statdesc}
\end{table}
}

\begin{table}[tbp]
{\normalsize 
\rotatebox{90}{
\begin{tabular}{|ll|l|l|ccc|ccc|ccc|ccc|}
\hline
& \multicolumn{3}{c|}{\textbf{FULL}} & \multicolumn{12}{|c|}{\textbf{ZONING}}
\\ 
& \multicolumn{3}{c|}{\textbf{SAMPLE}} & \multicolumn{12}{|c|}{\textbf{SCHEME}} \\ \hline
&  &  &  & \multicolumn{3}{|c|}{\textit{0}} & \multicolumn{3}{|c|}{\textit{5B}} & \multicolumn{3}{|c|}{\textit{ZRR\&5B}} & \multicolumn{3}{|c|}{\textit{ZRR}} \\ \hline
${\small t}$ & {\small Mean} & {\small M} & {\small Mode} & {\small Mean} & 
{\small M} & {\small Mode(\%)} & {\small Mean} & {\small M} & {\small Mode(\%)} & {\small Mean} & {\small M} & {\small Mode(\%)} & {\small Mean} & 
{\small M} & {\small Mode(\%)} \\ \hline
{\small 1994} & \multicolumn{1}{c|}{\small -0.06} & \multicolumn{1}{|c|}{\small 0} & \multicolumn{1}{|c|}{\small 0(26)} & {\small 0.21} & {\small 0}
& {\small 0(22)} & {\small \ -0.44} & {\small 0} & {\small 0(24)} & {\small -0.20} & {\small 0} & {\small 0(33)} & {\small -0.66} & {\small 0} & {\small 0(36)} \\ 
{\small 1995} & \multicolumn{1}{c|}{\small -0.27} & \multicolumn{1}{|c|}{\small 0} & \multicolumn{1}{|c|}{\small 0(20)} & {\small 0.06} & {\small 0}
& {\small 0(16)} & {\small -0.56} & {\small 0} & {\small 0(18)} & {\small -0.59} & {\small 0} & {\small 0(25)} & {\small -1.10} & {\small 0} & {\small 0(27)} \\ 
{\small 1996} & \multicolumn{1}{c|}{\small 1.08} & \multicolumn{1}{|c|}{\small 0} & \multicolumn{1}{|c|}{\small 0(16)} & {\small 1.69} & {\small 0}
& {\small 0(12)} & {\small 0.78} & {\small 0} & {\small 0(14)} & {\small 0.35} & {\small 0} & {\small 0(21)} & {\small -0.17} & {\small 0} & {\small 0(23)} \\ 
{\small 1997} & \multicolumn{1}{c|}{\small -1.01} & \multicolumn{1}{|c|}{\small 0} & \multicolumn{1}{|c|}{\small 0(13)} & {\small -0.99} & {\small 0}
& {\small 0(11)} & {\small -1.16} & {\small 0} & {\small 0(11)} & {\small -0.88} & {\small 0} & {\small 0(18)} & {\small -1.58} & {\small 0} & {\small 0(19)} \\ 
{\small 1998} & \multicolumn{1}{c|}{\small 0.76} & \multicolumn{1}{|c|}{\small 0} & \multicolumn{1}{|c|}{\small 0(12)} & {\small 1.27} & {\small 0}
& {\small 0(10)} & {\small 0.65} & {\small 0} & {\small 0(11)} & {\small 0.12} & {\small 0} & {\small 0(17)} & {\small -0.76} & {\small 0} & {\small 0(18)} \\ 
{\small 1999} & \multicolumn{1}{c|}{\small 4.58} & \multicolumn{1}{|c|}{\small 1} & \multicolumn{1}{|c|}{\small 0(11)} & {\small 5.30} & {\small 1}
& {\small 0(9)} & {\small 4.67} & {\small 1} & {\small 0(10)} & {\small 3.67}
& {\small 1} & {\small 0(15)} & {\small 0.06} & {\small 0} & {\small 0(15)}
\\ 
{\small 2000} & \multicolumn{1}{c|}{\small 5.28} & \multicolumn{1}{|c|}{\small 1} & \multicolumn{1}{|c|}{\small 0(11)} & {\small 6.49} & {\small 1}
& {\small 0(9)} & {\small 5.39} & {\small 1} & {\small 0(9)} & {\small 3.52}
& {\small 1} & {\small 0(15)} & {\small 0.29} & {\small 1} & {\small 0(14)}
\\ 
{\small 2001} & \multicolumn{1}{c|}{\small 8.72} & \multicolumn{1}{|c|}{\small 2} & \multicolumn{1}{|c|}{\small 0(10)} & {\small 10.75} & {\small 2}
& {\small 0(8)} & {\small 9.01} & {\small 2} & {\small 0(9)} & {\small 5.49}
& {\small 1} & {\small 0(13)} & {\small 2.45} & {\small 1} & {\small 0(14)}
\\ 
{\small 2002} & \multicolumn{1}{c|}{\small 12.88} & \multicolumn{1}{|c|}{\small 2} & \multicolumn{1}{|c|}{\small 0(9)} & {\small 16.32} & {\small 2}
& {\small 0(7)} & {\small 13.09} & {\small 3} & {\small 0(8)} & {\small 7.36}
& {\small 1} & {\small 0(13)} & {\small 4.66} & {\small 1} & {\small 0(13)}
\\ \hline
\end{tabular}} }
\caption{The evolution of employment overtime. Values refer to $\Delta $
Employment calculated as Employment(t)-Employment$(t_{0}).$ Time $t$ is
allowed to vary between 1994 to 2002 and $t_{0}$ is equal to 1993. $M$
indicates the median. The values between brackets indicate the relative
frequency in terms of percentage of the modal value.}
\label{table_desc2}
\end{table}
\end{center}

\bigskip {\normalsize 
\begin{table}[tbp]
{\normalsize 
\rotatebox{90}{
\begin{tabular}{||l|cc|cc|cc|cc|cc||}
\hline\hline
& \multicolumn{6}{|c|}{\textbf{Pre-program test}} & \multicolumn{4}{|c||}{\textbf{Post-program test}} \\ \hline
& $\alpha _{t}^{ZRR}$ & \texttt{p-value} & $\alpha _{t}^{ZRR}$ & 
\texttt{p-value} & $\alpha _{t}^{ZRR}$ & \texttt{p-value} & $\alpha
_{t}^{ZRR}$ & \texttt{p-value} & $\alpha _{t}^{ZRR}$ & \texttt{p-value}
\\ \hline
& \multicolumn{2}{|c|}{\small t=1994} & \multicolumn{2}{|c|}{\small t=1995}
& \multicolumn{2}{|c|}{\small t=1996} & \multicolumn{2}{|c|}{\small t=2001}
& \multicolumn{2}{|c||}{\small t=2002} \\ 
& \multicolumn{2}{|c|}{{\small t}$_{0}${\small =1993}} & 
\multicolumn{2}{|c|}{{\small t}$_{0}${\small =1993}} & \multicolumn{2}{|c|}{{\small t}$_{0}${\small =1993}} & \multicolumn{2}{|c|}{{\small t}$_{0}${\small =1993}} & \multicolumn{2}{|c||}{{\small t}$_{0}${\small =1993}} \\ 
\hline
\textbf{Before-After} &  &  &  &  &  &  &  &  &  &  \\ 
{\small No control variables} & {\small 0.271} & {\small 0.551} & {\small -0.113} & {\small 0.814} & {\small -0.459} & {\small 0.445} & {\small -3.523}
& {\small 3.245e-05} & {\small -5.775} & {\small 3.855e-08} \\ 
{\small Control variables }$(X_{i})$ & {\small 0.777} & {\small 0.207} & 
{\small 0.262} & {\small 0.703} & {\small 0.537} & {\small 0.475} & {\small 1.003} & {\small 0.358} & {\small 0.461} & {\small \ 0.719} \\ 
&  &  &  &  &  &  &  &  &  &  \\ \hline
& \multicolumn{2}{|c}{\small t=1995} & \multicolumn{2}{|c}{\small t=1996} & 
\multicolumn{2}{|c}{} & \multicolumn{2}{|c}{\small t=2001} & 
\multicolumn{2}{|c||}{\small t=2002} \\ 
& \multicolumn{2}{|c|}{{\small \ t}$_{0}${\small =1994}} & 
\multicolumn{2}{|c}{{\small t}$_{0}${\small =1994}} & \multicolumn{2}{|c}{}
& \multicolumn{2}{|c}{{\small t}$_{0}${\small =1994}} & 
\multicolumn{2}{|c||}{{\small t}$_{0}${\small =1994}} \\ \hline
\textbf{Before-After} &  &  &  &  &  &  &  &  &  &  \\ 
{\small No control variables} & {\small -0.462} & {\small 0.296} & {\small -0.728} & {\small 0.190} &  &  & {\small -3.707} & {\small 3.672e-06} & 
{\small -5.984} & {\small 2.625e-09} \\ 
{\small Control variables }$(X_{i})$ & {\small -0.541} & {\small 0.386} & 
{\small -0.200} & {\small 0.768} &  &  & {\small 0.301} & {\small 0.771} & 
{\small -0.237} & {\small 0.847} \\ 
&  &  &  &  &  &  &  &  &  &  \\ 
\textbf{Random growth} &  &  &  &  &  &  &  &  &  &  \\ 
{\small No control variables} & {\small -0.627} & {\small 0.311} & {\small -1.193} & {\small 0.233} &  &  & {\small -5.376} & {\small 0.0444} & {\small -7.819} & {\small 0.010} \\ 
{\small Control variables }$(X_{i})$ & {\small -1.114} & {\small 0.167} & 
{\small -1.338} & {\small 0.299} &  &  & {\small -3.681} & {\small 0.300} & 
{\small -4.787} & {\small 0.232} \\ 
& \multicolumn{1}{|l}{} & \multicolumn{1}{l|}{} & \multicolumn{1}{|l}{} & 
\multicolumn{1}{l|}{} & \multicolumn{1}{|l}{} & \multicolumn{1}{l|}{} & 
\multicolumn{1}{|l}{} & \multicolumn{1}{l|}{} & \multicolumn{1}{|l}{} & 
\multicolumn{1}{l||}{} \\ \hline\hline
\end{tabular}}  }
\caption{ Placebo tests}
\label{table_placebo}
\end{table}
}

\bigskip

{\normalsize 
\begin{table}[tbp]
{\normalsize 
\begin{tabular}{c|cccc|ccc}
\hline
& \multicolumn{4}{|c|}{CONTINUOUS\ PART} & \multicolumn{3}{|c}{DISCRETE\ PART
} \\ \hline
$t$ &  &  &  &  &  &  &  \\ \hline
& $\alpha _{t}^{ZRR}$ & $\alpha _{t}^{5B}$ & $\alpha _{t}^{ZRR\&5B}$ &  & $%
\delta \beta _{0t}^{ZRR}$ & $\delta \beta _{0t}^{5B}$ & $\delta \beta
_{0t}^{ZRR\&5B}$ \\ 
\begin{tabular}{l}
1996 \\ 
\end{tabular}
& 
\begin{tabular}{l}
-0.049 \\ 
(0.958)%
\end{tabular}
& 
\begin{tabular}{l}
0.114 \\ 
(0.868)%
\end{tabular}
& 
\begin{tabular}{l}
-0.010 \\ 
(0.990)%
\end{tabular}
&  & 
\begin{tabular}{c}
0.108 \\ 
(0.061)%
\end{tabular}
& 
\begin{tabular}{c}
-0.055 \\ 
(0.348)%
\end{tabular}
& 
\begin{tabular}{c}
0.053 \\ 
(0.334)%
\end{tabular}
\\ 
\begin{tabular}{l}
1997 \\ 
\end{tabular}
& 
\begin{tabular}{l}
-0.1418 \\ 
(0.882)%
\end{tabular}
& 
\begin{tabular}{l}
0.894 \\ 
(0.201)%
\end{tabular}
& 
\begin{tabular}{l}
0.764 \\ 
(0.415)%
\end{tabular}
&  & 
\begin{tabular}{c}
0.147 \\ 
(0.017)%
\end{tabular}
& 
\begin{tabular}{c}
-0.147 \\ 
(0.018)%
\end{tabular}
& 
\begin{tabular}{c}
-0.001 \\ 
(0.993)%
\end{tabular}
\\ 
\begin{tabular}{l}
1998 \\ 
\end{tabular}
& 
\begin{tabular}{l}
0.2119 \\ 
(0.8367)%
\end{tabular}
& 
\begin{tabular}{l}
0.864 \\ 
(0.251)%
\end{tabular}
& 
\begin{tabular}{l}
1.087 \\ 
(0.277)%
\end{tabular}
&  & 
\begin{tabular}{c}
0.0874 \\ 
(0.168)%
\end{tabular}
& 
\begin{tabular}{c}
-0.124 \\ 
(0.050)%
\end{tabular}
& 
\begin{tabular}{c}
-0.037 \\ 
(0.535)%
\end{tabular}
\\ 
\begin{tabular}{l}
1999 \\ 
\end{tabular}
& 
\begin{tabular}{l}
2.159 \\ 
(0.063)%
\end{tabular}
& 
\begin{tabular}{l}
1.378 \\ 
(0.100)%
\end{tabular}
& 
\begin{tabular}{l}
3.537 \\ 
(0.001)%
\end{tabular}
&  & 
\begin{tabular}{c}
0.047 \\ 
(0.463)%
\end{tabular}
& 
\begin{tabular}{c}
-0.131 \\ 
(0.044)%
\end{tabular}
& 
\begin{tabular}{c}
-0.083 \\ 
(0.178)%
\end{tabular}
\\ 
\begin{tabular}{l}
2000 \\ 
\end{tabular}
& 
\begin{tabular}{l}
1.372 \\ 
(0.258)%
\end{tabular}
& 
\begin{tabular}{l}
0.721 \\ 
(0.438)%
\end{tabular}
& 
\begin{tabular}{l}
2.381 \\ 
(0.044)%
\end{tabular}
&  & 
\begin{tabular}{c}
0.054 \\ 
(0.419)%
\end{tabular}
& 
\begin{tabular}{c}
-0.098 \\ 
(0.142)%
\end{tabular}
& 
\begin{tabular}{c}
-0.043 \\ 
(0.491)%
\end{tabular}
\\ 
\begin{tabular}{l}
2001 \\ 
\end{tabular}
& 
\begin{tabular}{l}
1.0862 \\ 
(0.418)%
\end{tabular}
& 
\begin{tabular}{l}
1.376 \\ 
(0.173)%
\end{tabular}
& 
\begin{tabular}{l}
2.454 \\ 
(0.059)%
\end{tabular}
&  & 
\begin{tabular}{c}
0.051 \\ 
(0.460)%
\end{tabular}
& 
\begin{tabular}{c}
-0.131 \\ 
(0.058)%
\end{tabular}
& 
\begin{tabular}{c}
-0.079 \\ 
(0.225)%
\end{tabular}
\\ 
\begin{tabular}{l}
2002 \\ 
\end{tabular}
& 
\begin{tabular}{l}
-0.174 \\ 
(0.912)%
\end{tabular}
& 
\begin{tabular}{l}
1.017 \\ 
(0.408)%
\end{tabular}
& 
\begin{tabular}{l}
1.279 \\ 
(0.406)%
\end{tabular}
&  & 
\begin{tabular}{c}
0.124 \\ 
(0.071)%
\end{tabular}
& 
\begin{tabular}{c}
-0.089 \\ 
(0.212)%
\end{tabular}
& 
\begin{tabular}{c}
0.034 \\ 
(0.594)%
\end{tabular}
\\ 
&  &  &  &  &  &  &  \\ \hline
& $g_{t}^{ZRR}$ & $g_{t}^{5B}$ & $g_{t}^{ZRR\&5B}$ &  &  &  &  \\ \hline
\begin{tabular}{l}
1996 \\ 
\end{tabular}
& 
\begin{tabular}{l}
10.666 \\ 
(3.33e-08)%
\end{tabular}
& 
\begin{tabular}{l}
7.766 \\ 
(4.72e-04)%
\end{tabular}
& 
\begin{tabular}{l}
10.574 \\ 
(1.67e-09)%
\end{tabular}
&  &  &  &  \\ 
\begin{tabular}{l}
1997 \\ 
\end{tabular}
& 
\begin{tabular}{l}
11.019 \\ 
(3.91e-08)%
\end{tabular}
& 
\begin{tabular}{l}
5.725 \\ 
(0.144)%
\end{tabular}
& 
\begin{tabular}{l}
11.034 \\ 
(2.83e-10)%
\end{tabular}
&  &  &  &  \\ 
\begin{tabular}{l}
1998 \\ 
\end{tabular}
& 
\begin{tabular}{l}
10.911 \\ 
(3.26e-10)%
\end{tabular}
& 
\begin{tabular}{l}
3.495 \\ 
(0.033)%
\end{tabular}
& 
\begin{tabular}{l}
10.960 \\ 
(2e-16)%
\end{tabular}
&  &  &  &  \\ 
\begin{tabular}{l}
1999 \\ 
\end{tabular}
& 
\begin{tabular}{l}
12.703 \\ 
(1.25e-15)%
\end{tabular}
& 
\begin{tabular}{l}
3.029 \\ 
(7.88e-04)%
\end{tabular}
& 
\begin{tabular}{l}
12.703 \\ 
(5.15e-15)%
\end{tabular}
& \multicolumn{1}{c|}{} &  &  &  \\ 
\begin{tabular}{l}
2000 \\ 
\end{tabular}
& 
\begin{tabular}{l}
13.195 \\ 
(3.24e-16)%
\end{tabular}
& 
\begin{tabular}{l}
7.750 \\ 
(6.66e-07)%
\end{tabular}
& 
\begin{tabular}{l}
13.232 \\ 
(1.36e-14)%
\end{tabular}
&  &  &  &  \\ 
\begin{tabular}{l}
2001 \\ 
\end{tabular}
& 
\begin{tabular}{l}
10.144 \\ 
(2e-16)%
\end{tabular}
& 
\begin{tabular}{l}
5.088 \\ 
(2.15e-04)%
\end{tabular}
& 
\begin{tabular}{l}
8.695 \\ 
(2e-16)%
\end{tabular}
&  &  &  &  \\ 
\begin{tabular}{l}
2002 \\ 
\end{tabular}
& 
\begin{tabular}{l}
7.977 \\ 
(5.43e-13)%
\end{tabular}
& 
\begin{tabular}{l}
7.842 \\ 
(4.73e-10)%
\end{tabular}
& 
\begin{tabular}{l}
8.285 \\ 
(2.47e-13)%
\end{tabular}
& \multicolumn{1}{c|}{} &  &  &  \\ \hline
\end{tabular}
}
\caption{Main results. For the continuous part, $\protect\alpha _{it}^{r}=%
\protect\alpha _{t}^{r}+g_{t}^{r}(SIZE,DENSITY)$ and non-isotropic tensor
product splines \citep{Wood2006low} are used for the bivariate functions $%
g_{t}^{r}(SIZE,DENSITY)$. For such nonparametric components: we report the
effective degrees of freedom with \texttt{p-values} in brackets. For the
parametric components of both continuous and discrete parts, $\protect\alpha %
_{t}^{r}$ and $\protect\delta \protect\beta _{0t}^{ZRR\_b}$, we report the
estimated coefficient with \texttt{p-values} in brackets.}
\label{table_main}
\end{table}
}

\bigskip

\clearpage 

\end{document}